\documentclass[prd,aps,twocolumn,superscriptaddress,nofootinbib,showpacs,10pt,preprintnumbers,longbibliography]{revtex4-1}

\usepackage{amsmath,amssymb,bm,multirow,float,bbold,hyperref}
\usepackage{graphicx}
\usepackage[american]{babel}
\usepackage{enumerate}
\usepackage{xcolor,colortbl}
\usepackage{mathtools}
\usepackage{booktabs}
\usepackage{subcaption}
\usepackage{soul}
\setstcolor{red}
\usepackage{makebox}
\allowdisplaybreaks

\newcommand\varpm{\mathbin{\vcenter{\hbox{%
  \oalign{\hfil$\scriptstyle\hspace{-0.2ex}+\hspace{-0.2ex}$\hfil\cr
          \noalign{\kern-.5ex}
          $\scriptscriptstyle({-})$\cr}%
}}}}

\newcommand\varmp{\mathbin{\vcenter{\hbox{%
  \oalign{\hfil$\scriptstyle\hspace{-0.2ex}-\hspace{-0.2ex}$\hfil\cr
          \noalign{\kern-.5ex}
          $\scriptscriptstyle({+})$\cr}%
}}}}

\begin{document}
\title{Classification of anomaly-free 2HDMs with a gauged $U(1)^\prime$ symmetry} 

\author{Franz Nottensteiner}
\affiliation{Department of Astronomy and Theoretical Physics, Lund University, SE-223 62 Lund, Sweden}

\author{Astrid Ordell}\thanks{astrid.ordell@thep.lu.se}
 \affiliation{Department of Astronomy and Theoretical Physics, Lund University, SE-223 62 Lund, Sweden}
 
\author{Roman Pasechnik}\thanks{roman.pasechnik@thep.lu.se}
\affiliation{Department of Astronomy and Theoretical Physics, Lund University, SE-223 62 Lund, Sweden}

\author{Hugo Ser\^odio}\thanks{hugo.serodio@thep.lu.se}
\affiliation{Department of Astronomy and Theoretical Physics, Lund University, SE-223 62 Lund, Sweden}

\date{\today}

\preprint{LU TP 19-42}

\begin{abstract}{
Two-Higgs-doublet models (2HDMs) with a flavored $U(1)^\prime$ gauge symmetry are a popular extension to the Standard Model (SM), yet they currently lack a complete survey. In this paper, we present a full classification of anomaly-free 2HDMs within the SM fermion content, resulting in a total of eleven distinct models. Four of these are relatively well-studied, while the rest either partially, or entirely, lack previous treatment. We study these textures under a variety of experimental bounds, focusing mainly on the previously unexplored models. This work is intended to act as a catalog to models worth considering in greater detail. 
}
\end{abstract}

\pacs{}

\maketitle

\section{Introduction}\label{s:intro}

With the discovery of the Higgs boson~\cite{Aad:2012tfa,Chatrchyan:2012xdj}, the Standard Model (SM) has reinforced itself as the most accurate description of high-energy physics.
While the SM has endured all kinds of experimental scrutiny, it still leaves room for sizable new physics (NP) effects. For the past decades, many extensions have been proposed. One of the most popular SM extensions is the two-Higgs-doublet model (2HDM)~\cite{Branco:2011iw,Ivanov:2017dad}, where a single additional copy of the Higgs doublet is added to the SM. Such a simple scenario introduces, nevertheless, new types of interactions; a general 2HDM implementation is plagued by tree-level flavor-changing neutral currents (FCNCs) mediated by the new scalars~\cite{Crivellin:2013wna}. To circumvent such a problem, the Yukawa textures are constrained with various additional symmetries. In the so-called natural flavor-conserving 2HDM implementations, an additional $\mathbb{Z}_2$ is added, leading to the complete absence of tree-level FCNCs~\cite{Glashow1977:M1,Paschos1977}. Larger symmetries can also be introduced, and a popular scenario is the one known as the Branco-Grimus-Lavoura (BGL) model~\cite{Branco:1996bq}, where a $\mathbb{Z}_{n\geq 3}$ symmetry is used. In the BGL, there are FCNCs present at tree level; however, due to the specific symmetry implementation, they turn out to be highly suppressed \textit{\`a la} minimal flavor violation~\cite{Buras:2000dm,DAmbrosio:2002vsn}. 

A complete survey of 2HDMs with an Abelian flavor symmetry, either discrete or continuous, has been done by Ferreira and Silva~\cite{Ferreira:2010ir} (later confirmed through different methods in Refs.~\cite{Serodio:2013gka,Ivanov:2013bka}). From the initially available $3^{18}$ model implementations, and under the imposition of physical constraints in the quark sector, the authors were able to reduce this number down to $246$ model candidates. In the vast majority of these implementations, the symmetry group is either the $U(1)$ or a $\mathbb{Z}_{n\geq 3}$, which from the scalar potential point of view is not distinguishable and will always lead to the presence of an accidental Goldstone boson~\cite{Ivanov:2011ae}. Since many of these implementations use the global version of the symmetry, a common practice is to softly break it in the scalar potential. Another way is to promote the symmetry to a local one, turning the extra Goldstone boson into the longitudinal component of the new gauge field, i.e.~the $Z^\prime$ boson. Some of the pioneering work on anomaly-free 2HDMs with a gauged Abelian symmetry can be found in Refs.~\cite{Davidson:1979wr,Davidson:1979wt,Davidson:1979nt}.  

In the present work, we look into the general classification and phenomenology of flavored $U(1)^\prime$ gauge symmetries in the context of 2HDMs. This was initiated as a master thesis project, and some of the details on the procedure can be found in Ref.~\cite{Franz:2017}.\footnote{Note that the author's family name has been changed from Teichmann to Nottensteiner.}  We find that, out of the $246$, only $11$ distinct models survive anomaly cancellation. Out of this small subset, five of the models can be found in the literature with partial or complete phenomenological studies. The remaining scenarios were virtually unexplored until now. All these findings assumed only the SM fermionic content. Neutrinos can get their mass by a large number of different mechanisms and, therefore, deserve a detailed study on their own.

The paper is organized as follows: In Sec.~\ref{s:U12HDM}, we introduce the framework, notation and the general procedure used for the classification. Section  \ref{s:AnomalyFree} contains a detailed implementation of all 11 viable models, while, in Sec.~\ref{s:pheno}, we present an analysis of phenomenological constraints of the new models. The results in this section serve as a first step in the phenomenological validation of such models with relatively light NP fields. We conclude and summarize our findings in Sec.~\ref{s:con}. Additional details on the scalar potential and the anomaly conditions can be found in the Appendixes.

\section{Abelian Flavor symmetries in 2HDMs}\label{s:U12HDM} 

\subsection{2HDM Yukawa sector}

In this work, we consider a 2HDM with an additional gauged $U(1)^\prime$ symmetry. There is always at least one Higgs doublet charged under $U(1)^\prime$; however, if the only scale entering into its spontaneous breaking is the electroweak (EW) scale, it becomes extensively challenging to accommodate a relatively heavy $Z^\prime$ gauge boson without deviating from the SM $Z$ currents.\footnote{Scenarios where the $Z^\prime$ are exceedingly light, due to a very small gauge coupling, can be an alternative. However, in the models presented here, the meson observables would anyway place very strong constraints on such scenarios (see Sec.~\ref{s:pheno}). We do not further explore this option.} The particle content, in the scalar sector, is thus enlarged by an additional scalar singlet $S$, charged under the new Abelian symmetry. 

We parameterize the scalar fields as

\begin{align}
\begin{split}
\label{eq:Scalars}
\Phi_a &= \frac{1}{\sqrt{2}}
\begin{pmatrix}
\sqrt{2}\phi^+_a\\
v_a\,e^{i\alpha_a}+R_a+iI_a
\end{pmatrix}\,,\\[0.8mm]
S &=\frac{1}{\sqrt{2}} \left(v_S\,e^{i\alpha_S}+\rho+i\eta\right)\,,
\end{split}
\end{align}
with $a=1,2$, $v_{a,S}$ the vacuum expectation values (VEVs) of the scalar fields and $\sqrt{v_1^2+v_2^2}=v\simeq 246\,\text{GeV}$. The presence of two Abelian symmetries allows us to rephase the scalar fields such that $\alpha_1=\alpha_2=0$ and $\alpha_S\neq0$ can be chosen without the loss of generality. This choice for the phases will be used throughout this paper. Note that the inclusion of the scalar singlet, even though essential for phenomenological tests, does not effect the model classification. For details on the scalar potential, see Appendix \ref{app:scalar}. 

The Yukawa interactions for quarks and charged leptons are, in the flavor basis, given by 

\begin{align}
\begin{split}
\label{eq:YukLag}
-\mathcal{L}_\mathrm{Yukawa}&= \overline{q_{\mathrm{L}}^0}\,\Gamma_a \Phi_a\, d^0_{\mathrm{R}} + \overline{q_{\mathrm{L}}^0}\,\Delta_a\tilde{\Phi}_a \, u^0_{\mathrm{R}}\\
&+ \overline{\ell_{\mathrm{L}}^0}\,\Pi_a\Phi_a\, e^0_{\mathrm{R}} + \mathrm{H.c.}\,,
\end{split}
\end{align}
with $\tilde{\Phi}_a=i\sigma_2{\Phi}_a^*$, where $\sigma_2$ is the Pauli matrix. As stated in the introduction, the fermionic content is the SM one with the fields defined in the generic flavor basis. After EW symmetry breaking, the mass matrices take the form

\begin{align}
\begin{split}
M_u&=\frac{1}{\sqrt{2}}\left(v_1\Delta_1+v_2\Delta_2\right)\,,\\
M_d&=\frac{1}{\sqrt{2}}\left(v_1\Gamma_1+v_2\Gamma_2\right)\,,\\
M_e&=\frac{1}{\sqrt{2}}\left(v_1\Pi_1+v_2\Pi_2\right)\,.
\end{split}
\end{align}
These matrices are then diagonalized via the unitary field transformations $f^0_\mathrm{L(R)}= U_\mathrm{L(R)}f_\mathrm{L(R)}$, such that
\begin{align} 
\label{eq:SVD}
U_\mathrm{fL}^\dagger M_f U_\mathrm{fR} = D_f\quad\text{with}\quad f=\{u,d,e\}\,,
\end{align}
where $D_f$ is a diagonal matrix with the ordered masses.

In the 2HDM there is a unique orthogonal combination, per fermionic sector, that can be defined and that carries all the information about flavor-changing iterations in the mass eigenbasis, namely,

\begin{align}
\label{eq:ortho}
\begin{split}
N_u&=\frac{1}{\sqrt{2}}U_\mathrm{uL}^\dagger\left(v_2\Delta_1-v_1\Delta_2\right)U_\mathrm{uR}\,,\\
N_d&=\frac{1}{\sqrt{2}}U_\mathrm{dL}^\dagger\left(v_2\Gamma_1-v_1\Gamma_2\right)U_\mathrm{dR}\,,\\
N_e&=\frac{1}{\sqrt{2}}U_\mathrm{eL}^\dagger\left(v_2\Pi_1-v_1\Pi_2\right)U_\mathrm{eR}\,.
\end{split}
\end{align}
Any off-diagonal component in Eq.~\eqref{eq:ortho} can be further emphasized by presenting it in the following form

\begin{align}
\label{eq:NdNu}
\begin{split}
N_u&={t}_\beta D_u-\left({t}_\beta+{t}_\beta^{-1}\right)\frac{v_2}{\sqrt{2}}\;U_\mathrm{uL}^\dagger\Delta_2U_\mathrm{uR}\,,\\
N_d&={t}_\beta D_d-\left({t}_\beta+{t}_\beta^{-1}\right)\frac{v_2}{\sqrt{2}} \;U_\mathrm{dL}^\dagger\Gamma_2U_\mathrm{dR}\,,\\
N_e&={t}_\beta D_e-\left({t}_\beta+{t}_\beta^{-1}\right)\frac{v_2}{\sqrt{2}} \;U_\mathrm{eL}^\dagger\Pi_2U_\mathrm{eR}\,,
\end{split}
\end{align}
with ${t}_\beta\equiv\tan \beta=v_2/v_1$, and where the latter term can be expressed in terms of projectors on the quark mass matrices. For further convenience, we can single out the sources of tree-level FCNCs mediated by scalar fields into the dimensionless quantities

\begin{align}
\label{eq:K}
\begin{split}
\mathcal{K}_u=&U_\mathrm{uL}^\dagger\Delta_2U_\mathrm{uR}\,,\quad \mathcal{K}_d = U_\mathrm{dL}^\dagger\Gamma_2U_\mathrm{dR}\,, \\
\mathcal{K}_e =& U_\mathrm{eL}^\dagger\Pi_2U_\mathrm{eR}\,.
\end{split}
\end{align}
The specific form for these quantities is texture dependent and will be specified for all models in Sec.~\ref{s:AnomalyFree}.

\subsection{The new gauge sector}
In this section, we introduce the relevant interactions of the new gauge field, in addition to discussing the role played by the scalar singlet. Under the corresponding Abelian symmetry, the field transformations are given by  

\begin{align}
\label{eq:transformation}
\begin{split}
&{q_\mathrm{L}^0}_j \rightarrow e^{i\alpha X_{q_j}}{q_\mathrm{L}^0}_j,\;{d_\mathrm{R}^0}_j \rightarrow e^{i\alpha X_{d_j}}{d_\mathrm{R}^0}_j , \;{u_\mathrm{R}^0}_j \rightarrow e^{i\alpha X_{u_j}}{u_\mathrm{R}^0}_j,\\
&{\ell_\mathrm{L}^0}_j \rightarrow e^{i\alpha X_{\ell_j}}{\ell_\mathrm{L}^0}_j,\;\;{e_\mathrm{R}^0}_j \rightarrow e^{i\alpha X_{e_j}}{e_\mathrm{R}^0}_j\\
&\Phi_a \rightarrow e^{i\alpha  X_{\Phi_a}} \Phi_a,\; S \rightarrow e^{i\alpha X_S} S\,.
\end{split}
\end{align}
The charges of the fields are, in general, flavor dependent and denoted by the label of the corresponding field. We will, throughout the paper, use a compact matrix notation for the charges. For example, for $q_\mathrm{L}$, 
\begin{align}
\label{eq:MatrixNotation}
\mathcal{X}^q\equiv \text{diag}(X_{q_1},X_{q_2},X_{q_3})\,,
\end{align}
and similarly for all the other fields. 

A convenient basis to write the relevant Lagrangian for the gauge sector is the would-be SM basis, i.e. the mass eigenbasis for the $Z$ gauge boson in the absence of mixing with the new $Z'$ boson. In such a basis, the fields and free parameters carry a hat. The neutral gauge interactions after EW symmetry breaking are then given by

\begin{align}
\label{eq:LZprime}
\begin{split}
\mathcal{L}_\mathrm{Z^\prime}=& -\frac{1}{4}\hat{Z}_{\mu\nu}\hat{Z}^{\mu\nu} -\frac{1}{4}\hat{Z}^\prime_{\mu\nu}\hat{Z}^{\prime \mu\nu}\\
&+\frac{1}{2}\hat{M}_\mathrm{Z}^2 \hat{Z}_\mu \hat{Z}^\mu + \frac{1}{2}\hat{M}_\mathrm{Z^\prime}^2 \hat{Z}^\prime_\mu \hat{Z}^{\prime\mu}+\delta \hat{M}_\mathrm{ZZ^\prime}^2 \hat{Z}_\mu \hat{Z}^{\prime\mu}\\
&-\hat{Z}^{(\prime)}_\mu\,\overline{\psi}\,\gamma^\mu \hat{Q}_\mathrm{Z^{(\prime)}}^{\psi}\, \psi\\
&-\hat{Z}^{(\prime)}_\mu\left(\partial_\mu\phi^\dagger\,  \hat{Q}_\mathrm{Z^{(\prime)}}^{\phi} \,\phi - \phi^\dagger \, \hat{Q}_\mathrm{Z^{(\prime)}}^{\phi}\, \partial_\mu\phi\right)\,,
\end{split}
\end{align}    
in addition to all quartic interactions, where $\psi=\{u^0_\mathrm{L/R},d^0_\mathrm{L/R},e^0_\mathrm{L/R},\nu^0_\mathrm{L}\}$ and $\phi=\{\phi_a^+,R_a,I_a,\rho,\eta\}$ and where we assume the gauge kinetic mixing to be zero. 

The first and second line of Eq.~\eqref{eq:LZprime} includes the mass terms for the $\hat{Z}-\hat{Z}^\prime$ system, which in the would-be SM basis are explicitly given by
\begin{align}
\label{eq:mass1}
\begin{split}
\hat{M}_\mathrm{Z}^2=&\frac{1}{4}\hat{f}^2v^2\,,\quad \hat{M}_\mathrm{Z^\prime}^2=g^{\prime 2}\left(v_a^2 X_{\Phi_a}^2+v_S^2 X_{S}^2\right)\,,\\
\delta \hat{M}_\mathrm{{Z}{Z}^\prime}^2=&-\frac{1}{2}\hat{f}g^\prime \left(v_a^2 X_{\Phi_a}\right)\,,
\end{split}
\end{align}
with $\hat{f}=\hat{e}/s_\theta c_\theta$, where $\hat{e}$ is the electromagnetic charge in the would-be SM basis. In the absence of mass mixing, i.e. $\delta \hat{M}^2=0$, the gauge boson $\hat{Z}$ is identified with the SM $Z$ boson and we can lose the hat notation. We would then have the $\overline{\text{MS}}$ weak mixing angle $s_\theta^2\simeq 0.231157(23)$, while the on-shell weak mixing angle, extracted by $\nu N$ experiments~\cite{Tanabashi:2018oca}, is given by $s_W^2\equiv 1-m_W^2/m_Z^2=s_\theta^2-\text{"EW-loop"}\simeq 0.2237(9)$. For a nonzero mass mixing, the value of $s_\theta$ gets modified~\cite{Babu:1997st} (see Sec.~\ref{s:pheno} for more details).

In the most general scenario, the mixing between $\hat{Z}$ and $\hat{Z}^\prime$ is removed through the orthogonal field transformation

\begin{align}
\label{eq:orthogoRZ}
\begin{pmatrix}
\hat{Z}_\mu\\
\hat{Z}^\prime_\mu
\end{pmatrix}=
\begin{pmatrix}
c_{M}&s_M\\
-s_M&c_M
\end{pmatrix}
\begin{pmatrix}
Z_\mu\\
Z^\prime_\mu
\end{pmatrix}\,,\quad s_M\equiv \sin(\theta_M)\,,
\end{align}
with
\begin{align}
\label{eq:tanM}
\tan(2\theta_M)=\frac{2\delta \hat{M}_\mathrm{ZZ^\prime}^2}{\hat{M}_\mathrm{Z^\prime}^2-\hat{M}_\mathrm{Z}^2}\,,
\end{align}
such that, in the mass eigenbasis for the gauge fields, the masses of $Z$ and $Z'$ are given by 

\begin{align}
\label{eq:mass2}
\begin{split}
m_{Z^{(\prime)}}^2&=\hat{M}_{Z^{(\prime)}}^2\varmp t_M\delta \hat{M}^2_{ZZ'}\,.
\end{split}
\end{align}

In the final two lines of Eq.~\eqref{eq:LZprime} we have interactions between the massive neutral gauge bosons and the fermionic and bosonic matter, respectively. The couplings are matrices in flavor space, explicitly given by
\begin{align}
\label{eq:QZZp}
\begin{split}
\hat{Q}_\mathrm{Z}^{i}&=\hat{f}(T^i_3-s_\theta^2 Q^i)\,,\quad
\hat{Q}_\mathrm{Z^\prime}^{i}= g^\prime \mathcal{X}^i\,,
\end{split}
\end{align}
with $i=u_\mathrm{L,R},\;d_\mathrm{L,R},\;\ell_\mathrm{L},\;e_\mathrm{L}$. In the mass eigenbasis for the gauge fields, we then have that

\begin{align}
\label{eq:QZZp_mass}
\begin{split}
Q_\mathrm{Z}^{i}=&c_M \hat{Q}_\mathrm{Z}^{i}-s_M \hat{Q}_\mathrm{Z^\prime}^{i}\,,\quad
Q_\mathrm{Z^\prime}^{i}=s_M \hat{Q}_\mathrm{Z}^{i}+c_M \hat{Q}_\mathrm{Z^\prime}^{i}\,,
\end{split}
\end{align}
where $T_3$ is the isospin and $Q$ the electric charge, and with $Q\equiv T_3 + Y$, where $Y$ is the hypercharge.

Note that, in the mass eigenbasis for the matter fields, $\hat{Q}_\mathrm{Z}^{i}$ retain its form, while $\hat{Q}_\mathrm{Z^\prime}^{i}$ acquires a generic flavor structure, i.e.

\begin{align}
\label{eq:epsi_Zprime}
\begin{split}
\hat{Q}_\mathrm{Z^\prime}^{i}\longrightarrow \hat{Q}_\mathrm{Z^\prime}^{i}= g^\prime U_i^\dagger \mathcal{X}^i U_i\equiv g^\prime \Xi_i.
\end{split}
\end{align}
In accordance with the scalar sector, all sources of tree-level FCNCs mediated by the neutral gauge bosons, are encoded in the dimensionless quantity $\Xi_i$. Further details on the specific form of these couplings are given for all models in Sec.~\ref{s:AnomalyFree}.

\subsection{General procedure for anomaly-free solutions}\label{subsec:conditions}

In 2011, Ferreira and Silva classified all possible implementations of a global Abelian symmetry in the quark sector of a 2HDM \cite{Ferreira:2010ir}. In short, they used that a flavored symmetry transformation, such as the one defined in Eq.~\eqref{eq:transformation}, imposes constraints on the Yukawa couplings. In order for the Yukawa Lagrangian, in Eq.~\eqref{eq:YukLag}, to remain invariant under such field transformations, the following constraints have to be fulfilled (for the quark sector) 

\begin{align}
\label{eq:SymmetryTextures}
\begin{split}
(\Gamma_a)_{ij} &= e^{i\theta(X_{q_i}-X_{d_j}- X_{\Phi_a})}(\Gamma_a)_{ij}\,,
\\
(\Delta_a)_{ij} &= e^{i\theta(X_{q_i}-X_{u_j}+ X_{\Phi_a})}(\Delta_a)_{ij}\,, 
\end{split}
\end{align}
such that

\begin{align}
\label{eq:ResTextures}
\begin{split}
(\Gamma_a)_{ij}&=\mathrm{any\;\;\;\;if}\;\;X_{q_i}-X_{d_j}= X_{\Phi_a}\,, 
\\
(\Gamma_a)_{ij}&=0\;\;\;\;\;\;\;\;\mathrm{if}\;\;X_{q_i}-X_{d_j}\neq X_{\Phi_a}\,, 
\end{split}
\end{align}
and similarly for the up sector, with $a=1,2$, $j=1,2,3$, and with no summation over repeated indices.\footnote{The case of a possible discrete Abelian symmetry was also tackled in~\cite{Ferreira:2010ir}. However, those cases are of no relevance in our study.} On top of these constraints, one can add three extra physical requirements:

\begin{itemize}
	\item[(i)] no massless up-type quarks, i.e.~$\text{det} \hspace{0.5mm}M_u\neq 0$\hspace{0.5mm}; 
	\item[(ii)] no massless down-type quarks, i.e.~$\text{det}\hspace{0.5mm} M_d\neq 0$\hspace{0.5mm};
	\item[(iii)] possibility for Dirac-type $CP$ violation at tree level, i.e.~$\text{det} [M_uM_u^\dagger,M_dM_d^\dagger]\neq 0$\hspace{0.5mm}.
\end{itemize}
In Ref.~\cite{Ferreira:2010ir}, a weaker formulation of (iii) was used, demanding the Cabibbo-Kobayashi-Maskawa (CKM) mixing matrix to be non-block diagonal.\footnote{For the models obtained, there is no real difference between the two formulations, but the latter has a more straightforward implementation when dealing with generic textures.}
From these simple requirements, in addition to excluding any models equivalent up to permutations,\footnote{Any transformation that preserves the flavor symmetry, i.e. any simultaneous permutation of rows or independent permutation of columns, will result in an equivalent model. The row (column) permutations simply alter the constraints in such a way that the corresponding left (right) charges are exchanged, resulting in the procedure merely amounting to a relabelling of flavor indices.} the number of viable texture combinations was reduced down from $3^{18}$ to $246$. 

In contrast to Ferreira and Silva, we now wish to classify all possible implementations of a \emph{gauged} Abelian symmetry in the 2HDM, in addition to including not only the SM quarks but also the SM charged leptons. With the Abelian symmetry now being gauged, we must also make sure that the solutions are anomaly-free. The six anomaly conditions which do not cancel trivially are presented in Appendix \ref{app:anomaliesCond} and involve

\begin{align}
\label{eq:ano1}
\begin{split}
&[{SU}(2)_\mathrm{L}]^2{U}(1)^\prime, \;\;\;[{SU}(3)_\mathrm{C}]^2{U}(1)^\prime, 
\\
&[{U}(1)_\mathrm{Y}]^2{U}(1)^\prime,\;\;\;
{U}(1)_\mathrm{Y}[{U}(1)^\prime]^2,\\ &[{U}(1)^\prime]^3,\;\;\;[\mathrm{Gravity}]^2{U}(1)^\prime\,.
\end{split}
\end{align}
There has been plenty of activity over the past decades on finding efficient ways of extracting generic solutions of such a system of equations~\cite{Batra:2005rh,Allanach:2018vjg,Rathsman:2019wyk,Costa:2019zzy}.\footnote{There are different approaches where the anomaly constraints can be relaxed. The so-called Green-Schwarz mechanism~\cite{Green:1984sg,Green:1984qs,Green:1984bx} or effective anomalous $U(1)^\prime$ scenarios~\cite{Preskill:1990fr,Coriano:2007fw,Coriano:2007xg,Ekstedt:2017tbo}, are some of the popular ones. We shall not pursue this line further and consider only models that cancel all the anomaly equations.}

To find all valid models, we use, as a starting point, the subset of textures in Ref.~\cite{Ferreira:2010ir} corresponding to a continuous symmetry. The textures are then, in correspondence with the anomaly conditions, converted into constraints for the charges just like in Eq.~\eqref{eq:ResTextures}. When including the charged leptons, there is no need to perform the whole procedure in Ref.~\cite{Ferreira:2010ir} once more, as the lepton textures are highly constrained by the anomaly equations. Instead, we need only to impose the additional physical requirement
\begin{itemize}
\item[(iv)] no massless charged leptons, i.e. $\text{det}\hspace{0.5mm}M_e\neq 0$.
\end{itemize}

For a 2HDM, this additional requirement can be fulfilled only when the combined texture of $\Pi_1$ and $\Pi_2$ have at least one nonzero entry in each row and each column. There are, hence, six possible minimal combined textures, but, since they are all equivalent up to permutations, it is of interest to consider only one of them, e.g.~the diagonal one
\begin{align}
\label{eq:leptonTexture1}
\begin{split}
(\mathrm{I})&\;\;\;\;\;\Pi_1:
\begin{pmatrix}
 \times &   \makebox[\widthof{$\times$}][c]{}&  \makebox[\widthof{$\times$}][c]{}
\\
  \makebox[\widthof{$\times$}][c]{} &  \times&   \makebox[\widthof{$\times$}][c]{}
\\
 \makebox[\widthof{$\times$}][c]{} &   \makebox[\widthof{$\times$}][c]{}&   \times
\end{pmatrix}
\hspace{3mm}
\Pi_2:
\begin{pmatrix}
 \makebox[\widthof{$\times$}][c]{}& \makebox[\widthof{$\times$}][c]{}& \makebox[\widthof{$\times$}][c]{}
\\
 \makebox[\widthof{$\times$}][c]{}& \makebox[\widthof{$\times$}][c]{}& \makebox[\widthof{$\times$}][c]{}
\\
\makebox[\widthof{$\times$}][c]{} & \makebox[\widthof{$\times$}][c]{} & \makebox[\widthof{$\times$}][c]{}
\end{pmatrix},
\\
(\mathrm{II})&\;\;\;\;\;\Pi_1:
\begin{pmatrix}
 \times &   \makebox[\widthof{$\times$}][c]{}&  \makebox[\widthof{$\times$}][c]{}
\\
  \makebox[\widthof{$\times$}][c]{} &  \times&   \makebox[\widthof{$\times$}][c]{}
\\
 \makebox[\widthof{$\times$}][c]{} &   \makebox[\widthof{$\times$}][c]{}&   \makebox[\widthof{$\times$}][c]{}
\end{pmatrix}
\hspace{3mm}
\Pi_2:
\begin{pmatrix}
 \makebox[\widthof{$\times$}][c]{}& \makebox[\widthof{$\times$}][c]{}& \makebox[\widthof{$\times$}][c]{}
\\
 \makebox[\widthof{$\times$}][c]{}& \makebox[\widthof{$\times$}][c]{}& \makebox[\widthof{$\times$}][c]{}
\\
\makebox[\widthof{$\times$}][c]{} & \makebox[\widthof{$\times$}][c]{} &  \times
\end{pmatrix},
\\
(\mathrm{III})&\;\;\;\;\;\Pi_1:
\begin{pmatrix}
 \times &   \makebox[\widthof{$\times$}][c]{}&  \makebox[\widthof{$\times$}][c]{}
\\
  \makebox[\widthof{$\times$}][c]{} &   \makebox[\widthof{$\times$}][c]{}&   \makebox[\widthof{$\times$}][c]{}
\\
 \makebox[\widthof{$\times$}][c]{} &   \makebox[\widthof{$\times$}][c]{}&   \makebox[\widthof{$\times$}][c]{}
\end{pmatrix}
\hspace{3mm}
\Pi_2:
\begin{pmatrix}
 \makebox[\widthof{$\times$}][c]{}& \makebox[\widthof{$\times$}][c]{}& \makebox[\widthof{$\times$}][c]{}
\\
 \makebox[\widthof{$\times$}][c]{}& \times& \makebox[\widthof{$\times$}][c]{}
\\
\makebox[\widthof{$\times$}][c]{} & \makebox[\widthof{$\times$}][c]{} &  \times
\end{pmatrix},
\\
(\mathrm{IV})&\;\;\;\;\;\Pi_1:
\begin{pmatrix}
 \makebox[\widthof{$\times$}][c]{} &   \makebox[\widthof{$\times$}][c]{}&  \makebox[\widthof{$\times$}][c]{}
\\
  \makebox[\widthof{$\times$}][c]{} &   \makebox[\widthof{$\times$}][c]{}&   \makebox[\widthof{$\times$}][c]{}
\\
 \makebox[\widthof{$\times$}][c]{} &   \makebox[\widthof{$\times$}][c]{}&   \makebox[\widthof{$\times$}][c]{}
\end{pmatrix}
\hspace{3mm}
\Pi_2:
\begin{pmatrix}
 \times &   \makebox[\widthof{$\times$}][c]{}&  \makebox[\widthof{$\times$}][c]{}
\\
  \makebox[\widthof{$\times$}][c]{} &  \times&   \makebox[\widthof{$\times$}][c]{}
\\
 \makebox[\widthof{$\times$}][c]{} &   \makebox[\widthof{$\times$}][c]{}&   \times
\end{pmatrix},
\end{split}
\end{align}
corresponding to the constraints
 
\begin{itemize}
\item[(I)] \;$\;\;X_{\ell_i}-X_{e_i}=X_{\Phi_1}$\hspace{0.5mm},
\item[(II)] \;$\;\; X_{\ell_{1(2)}}-X_{e_{1(2)}}=X_{\Phi_1}$\quad and \quad$X_{\ell_{3}}-X_{e_{3}}=X_{\Phi_2}$\hspace{0.5mm},
\item[(III)] \;$\;\; X_{\ell_{1}}-X_{e_{1}}=X_{\Phi_1}$\quad and \quad$X_{\ell_{2(3)}}-X_{e_{2(3)}}=X_{\Phi_2}$\hspace{0.5mm},
\item[(IV)] \;$\;\;X_{\ell_i}-X_{e_i}=X_{\Phi_2}$\hspace{0.5mm}.
\end{itemize}

Note that these conditions are sufficient for generating any allowed lepton texture, and not just the minimal ones in $(\mathrm{I})-(\mathrm{IV})$, as constraints on the form $X_{\ell_{i}}-X_{e_{j}}\neq X_{\Phi_a}$ are not included. In our case, however, the textures in $(\mathrm{I})-(\mathrm{IV})$ are, in fact, the only ones allowed, as the anomaly conditions are highly constraining. The only exception to this is a special case of the models presented in Sec.~\ref{subsec:valid}, for which all leptons, and the Higgs doublet that they couple to, have zero $U(1)'$ charge. For such a scenario, the leptons are free to have any texture, provided that the physical condition in $(\mathrm{iv})$ is still fulfilled.

If a solution exists where all constraints are met, while simultaneously all charges are rational numbers and the quarks textures are not destroyed, the model is labeled as valid.\footnote{For an explicit example of a valid and invalid model, see~\cite{Franz:2017}} Out of the 246 models in Ref.~\cite{Ferreira:2010ir}, there are 116 of them corresponding to continuous symmetries, out of which only 11 nondegenerate models survive these constraints. Here, we emphasize that we do not scan to find viable solutions; we loop over every possible combination of constraints and solve the system for each instance. As such, we are guaranteed to find all solutions. 

To identify degenerate solutions, we consider all possible permutations of the textures, i.e.~every possible combination of $i,j,k$ for the transformation

\begin{equation}
\Gamma'_{1,2}=\mathcal{P}_i^{\mathrm{T}} \Gamma_{1,2} \mathcal{P}_j,\;\;\Delta'_{1,2}= \mathcal{P}_i^{\mathrm{T}} \Delta_{1,2} \mathcal{P}_k,
\end{equation}
where $\mathcal{P}$ is the three-dimensional representation of the permutation group $S_3$, such that $i,j,k$ ranges from one to six. The permutations on the left are shared by both quark sectors, while the permutations on the right are independent. With this corresponding to an ordered sampling with replacement, there is a total of $6^3=216$ possible permutations for each model. 

\section{Anomaly-Free Models}\label{s:AnomalyFree}  

Up to permutations, there are a total of seven valid models. Four of these come in two editions, $a$ and $b$, where $b$ has the flipped texture with respect to $a$, i.e. {\small$\Gamma_1\leftrightarrow\Delta_2$, $\Gamma_2\hspace{-0.5mm}\leftrightarrow\hspace{-0.5mm}\Delta_1$}. For each model, we present the allowed textures and some analytical predictions, with the corresponding charges specified in Table \ref{tab:charges}. An overview on which of the models that have been studied previously, in either its global or gauged form, can be found in Table \ref{tab:modelS}. 

As a first validity check for the less-studied models, this section includes a parameter scan for fitting the quark masses and mixings (MMs). The best-fit points then serve as input values when accounting for all the remaining observables in Sec.~\ref{s:pheno}. 

For all models, except M5 and M6, the free parameters in the Yukawa sector are

\begin{itemize}
	\item Yukawa couplings modulus $\in [10^{-5},5]$;
	
	\item Yukawa couplings argument $\in [10^{-10},2\pi[$;
	
	\item doublet fields VEVs, i.e. $\tan\beta\in [10^{-3},10^3]$.
\end{itemize}

The scan was performed by giving, as input, 300 different $\tan\beta$ values evenly distributed in logarithmic scale. For the models M5 and M6, on the other hand, it is more convenient to parameterize the Yukawa couplings with the physical quark masses and CKM mixing angles (see the respective model subsection for the explicit parameterization). For the fermions masses and CKM mixing matrix data, we have used Ref.~\cite{Tanabashi:2018oca}. The charged lepton mass matrix is diagonal in all models and, therefore, the extraction of their Yukawa couplings is straightforward.

\begin{table}
\begin{tabular}{lll}
\hline\hline
 \rule{0pt}{2.8ex}{\small{Model}}\;\;\;\;\;\;\;\;\;\;\;\; \;\;\;\;\;&  {\small{Global $U(1)^\prime$}} \;\;\;\;\;\;\;\;\;\;\;\;\;\;\;\;\;  &  {\small{Gauged $U(1)^\prime$}} \\
\hline
 \rule{0pt}{2.8ex}{\small{{M1}}}  & \cite{Glashow1977:M1,Haber:1978jt} &  \cite{Crivellin:2015mga,Camargo:2018klg,Campos:2017dgc,Camargo:2019ukv}   \\
 \rule{0pt}{2.8ex}{\small{{M2}}}  & -     &  -        \\
 \rule{0pt}{2.8ex}{\small{{M3a,\hspace{0.5mm}b}}}  &-     &  \cite{Celis:2016ayl,Crivellin:2015lwa,Bian:2017xzg,Bian:2017rpg,Ko:2019tts}     \\
 \rule{0pt}{2.8ex}{\small{{M4a,\hspace{0.5mm}b}}}  &-     &  -     \\
 \rule{0pt}{2.8ex}{\small{{M5}}} &  \cite{Alves:2018kjr}    &  \cite{Nomura:2017ezy}     \\
 \rule{0pt}{2.8ex}{\small{{M6a,\hspace{0.5mm}b}}}  &  \cite{Branco:1996bq,Botella:2014ska}    & \cite{Celis:2015ara}     \\
  \rule{0pt}{2.8ex}{\small{{M7a,\hspace{0.5mm}b}}}  & -     &  -   \\
\hline\hline
\end{tabular}
\caption{An overview of the valid models, where previous studies are indicated by the corresponding reference.}
\label{tab:modelS}
\end{table}

\subsection{Leptons that couple exclusively to one Higgs}\label{subsec:valid}

In this section, we present all models fulfilling the constraints specified in Sec.~\ref{subsec:conditions}, while simultaneously having the leptons coupling to only one of the Higgs doublets. In all these models, there exists a particular charge assignment for which one of the scalar doublets, as well as all charged leptons, is uncharged under $U(1)^\prime$. In this special case, all models can fulfill the anomaly constraints even without the introduction of charged leptons.

\subsubsection{Model M1}

Let us begin with the one model that is naturally flavor conserving, namely, 

\begin{align}
\label{eq:L8texture}
\begin{split}
\Gamma_1&:
\begin{pmatrix}
 \times &  \times&  \times
\\
 \times &  \times&  \times
\\
 \times &  \times&  \times
\end{pmatrix}
\hspace{2mm}
\Gamma_2:
\begin{pmatrix}
 \makebox[\widthof{$\times$}][c]{0}& \makebox[\widthof{$\times$}][c]{0}& \makebox[\widthof{$\times$}][c]{0}
\\
 \makebox[\widthof{$\times$}][c]{0}& \makebox[\widthof{$\times$}][c]{0}& \makebox[\widthof{$\times$}][c]{0}
\\
\makebox[\widthof{$\times$}][c]{0} & \makebox[\widthof{$\times$}][c]{0} & \makebox[\widthof{$\times$}][c]{0}
\end{pmatrix}
\hspace{2mm}
\\
\Delta_1&:
\begin{pmatrix}
 \times &  \times&  \times
\\
 \times &  \times&  \times
\\
 \times &  \times&  \times
\end{pmatrix}
\hspace{1,5mm}
\Delta_2:
\begin{pmatrix}
\makebox[\widthof{$\times$}][c]{0} &\makebox[\widthof{$\times$}][c]{0} & \makebox[\widthof{$\times$}][c]{0}
\\
 \makebox[\widthof{$\times$}][c]{0}& \makebox[\widthof{$\times$}][c]{0}& \makebox[\widthof{$\times$}][c]{0}
\\
 \makebox[\widthof{$\times$}][c]{0}& \makebox[\widthof{$\times$}][c]{0}& \makebox[\widthof{$\times$}][c]{0}
\end{pmatrix},
\end{split}
\end{align}
with the corresponding charges specified in Table \ref{tab:charges}. As is familiar, this model has no tree-level FCNCs, since $\mathcal{K}_{d,u}=0$ and $\Xi_i$ are diagonal.

For the particular choice of $y=3x$, the $U(1)^\prime$ symmetry places no constraints on the Yukawa quark textures, and we instead end up with the most general 2HDM.  This trivially allowed texture is not considered further in this work. 




\begin{table*}[t]
		\centering
		{\small
			\begin{tabular}{p{0.0002\linewidth} >{\raggedleft}p{0.15\linewidth}>{\raggedleft}p{0.14\linewidth} >{\raggedleft}p{0.14\linewidth} >{\raggedleft}p{0.15\linewidth} >{\raggedleft}p{0.13\linewidth} >{\raggedleft}p{0.13\linewidth} r}
				Model&$X_{q_L}$&$X_{u_R}$&$X_{d_R}$&$X_{\ell_L}$&$X_{e_R}$&$X_{\Phi}$&Cond.\\
				\hline\hline\\
				M1&$x\left[\begin{array}{c}1\\1\\1\end{array}\right]$&$4x\left[\begin{array}{c}1\\1\\1\end{array}\right]$
				&$-2x\left[\begin{array}{c}1\\1\\1\end{array}\right]$&$\left[\begin{array}{c}-3x\\3x+z\\-9x-z\end{array}\right]$
				&$\left[\begin{array}{c}-6x\\z\\-12x-z\end{array}\right]$&$\left[\begin{array}{c}3x\\y\end{array}\right]$&$y\neq 3x$\\\\
				\hline\\
				M2&$\left[\begin{array}{c}x\\y\\-x+2y\end{array}\right]$&$\left[\begin{array}{c}x+3y\\4y\\-x+5y\end{array}\right]$
				&$\left[\begin{array}{c}x-3y\\-2y\\-x-y\end{array}\right]$
				&$\left[\begin{array}{c}-3y\\3y+z\\-9y-z\end{array}\right]$&$\;\;\;\;\;\left[\begin{array}{c}-6y\\z\\-12y-z\end{array}\right]$
				&$\left[\begin{array}{c}3y\\x+2y\end{array}\right]$&$x\neq y$\\\\
				\hline\\
				M3a&$\left[\begin{array}{c}x\\x\\y\end{array}\right]$&$\left[\begin{array}{c}3x+y\\3x+y\\2(x+y)\end{array}\right]$
				&$-\left[\begin{array}{c}x+y\\x+y\\2x\end{array}\right]$&$\left[\begin{array}{c}-2x-y\\2x+y+z\\-6x-3y-z\end{array}\right]$
				&$\left[\begin{array}{c}-4x-2y\\z\\-8x-4y-z\end{array}\right]$&$\left[\begin{array}{c}2x+y\\x+2y\end{array}\right]$&$x\neq y$\\\\
				M3b&$--$&$--$&$--$&$--$&$--$&$\left[\begin{array}{c}3x\\2x+y\end{array}\right]$&$x\neq y$\\\\
				\hline\\
				M4a&$\left[\begin{array}{c}x\\y\\-x+2y\end{array}\right]$&$\dfrac{1}{3}\left[\begin{array}{c}5x+7y\\-4x+16y\\-x+13y\end{array}\right]$
				&$\dfrac{1}{3}\left[\begin{array}{c}x-7y\\x-7y\\-2x-4y\end{array}\right]$&$\dfrac{1}{3}\left[\begin{array}{c}x-10y\\8x-17y\\-9x\end{array}\right]$
				&$\dfrac{1}{3}\left[\begin{array}{c}2x-20y\\9x-27y\\-11x-7y\end{array}\right]$&$\dfrac{1}{3}\left[\begin{array}{c}-x+10y\\2x+7y\end{array}\right]$&$x\neq y$\\\\
				M4b&$--$&$\dfrac{1}{3}\left[\begin{array}{c}x+11y\\x+11y\\-2x+14y\end{array}\right]$
				&$\dfrac{1}{3}\left[\begin{array}{c}5x-11y\\-4x-2y\\-x-5y\end{array}\right]$&$\dfrac{1}{3}\left[\begin{array}{c}-9x\\10x-19y\\-x-8y\end{array}\right]$
				&$\;\;\dfrac{1}{3}\left[\begin{array}{c}-7x-11y\\9x-27y\\-2x-16y\end{array}\right]$&$\;\;\;\dfrac{1}{3}\left[\begin{array}{c}-2x+11y\\x+8y\end{array}\right]$&$x\neq y$\\\\
				\hline\\
				M5&$x\left[\begin{array}{c}1\\1\\1\end{array}\right]$&$\left[\begin{array}{c}2x-y\\2x-y\\8x+2y\end{array}\right]$
				&$\left[\begin{array}{c}y\\y\\-6x-2y\end{array}\right]$&$\left[\begin{array}{c}-x+y\\-5x-y\\-3x\end{array}\right]$
				&$2\left[\begin{array}{c}-x+y\\-3x\\-5x-y\end{array}\right]$&$\left[\begin{array}{c}x-y\\7x+2y\end{array}\right]$&$y\neq -2x$\\\\
				\hline\\
				M6a&$\left[\begin{array}{c}x\\x\\y\end{array}\right]$&$\dfrac{2}{3}\left[\begin{array}{c}5x+y\\5x+y\\2x+4y\end{array}\right]$
				&$-\dfrac{4x+2y}{3}\left[\begin{array}{c}1\\1\\1\end{array}\right]$&$\dfrac{1}{3}\left[\begin{array}{c}-7x-2y\\-11x+2y\\-9y\end{array}\right]$
				&$-\dfrac{2}{3}\left[\begin{array}{c}7x+2y\\9x\\2x+7y\end{array}\right]$&$\dfrac{1}{3}\left[\begin{array}{c}7x+2y\\4x+5y\end{array}\right]$&$x\neq y$\\\\
				M6b&$--$&$\dfrac{8x+4y}{3}\left[\begin{array}{c}1\\1\\1\end{array}\right]$
				&$-\dfrac{2}{3}\left[\begin{array}{c}x+2y\\x+2y\\4x-y\end{array}\right]$&$\dfrac{1}{3}\left[\begin{array}{c}-9y\\-13x+4y\\-5x-4y\end{array}\right]$
				&$-\dfrac{2}{3}\left[\begin{array}{c}4x+5y\\9x\\5x+4y\end{array}\right]$&$\dfrac{1}{3}\left[\begin{array}{c}8x+y\\5x+4y\end{array}\right]$&$x\neq y$\\\\
				\hline\\
				M7a&$\left[\begin{array}{c}x\\x\\y\end{array}\right]$&$\dfrac{1}{3}\left[\begin{array}{c}7x+5y\\7x+5y\\10x+2y\end{array}\right]$
				&$-\dfrac{1}{3}\left[\begin{array}{c}4x+2y\\x+5y\\7x-y\end{array}\right]$&$-\dfrac{1}{3}\left[\begin{array}{c}2x+7y\\7x+2y\\9x\end{array}\right]$
				&$-\dfrac{1}{3}\left[\begin{array}{c}9(x+y)\\14x+4y\\13x+5y\end{array}\right]$&$\dfrac{1}{3}\left[\begin{array}{c}7x+2y\\4x+5y\end{array}\right]$&$x\neq y$\\\\
				M7b&$--$&$\dfrac{1}{3}\left[\begin{array}{c}8x+4y\\11x+y\\5x+7y\end{array}\right]$
				&$-\dfrac{1}{3}\left[\begin{array}{c}5x+y\\5x+y\\2x+4y\end{array}\right]$&$-\dfrac{1}{3}\left[\begin{array}{c}9x\\4x+5y\\5x+4y\end{array}\right]$
				&$-\dfrac{1}{3}\left[\begin{array}{c}17x+y\\9(x+y)\\10x+8y\end{array}\right]$&$\dfrac{1}{3}\left[\begin{array}{c}8x+y\\5x+4y\end{array}\right]$&$x\neq y$\\\\
				\hline\hline
			\end{tabular}
		}
		\caption{Allowed charges for the various models, where the condition in the rightmost column is required in order for the textures to be conserved. The double dashes are used to indicate entries with identical charges to the model above. }
		\label{tab:charges}
	\end{table*}

\subsubsection{Model M2}

Next, we have the textures

\begin{align}
\label{eq:textureM2}
\begin{split}
\Gamma_1&:
\begin{pmatrix}
 \times & \makebox[\widthof{$\times$}][c]{0} & \makebox[\widthof{$\times$}][c]{0}
\\
\makebox[\widthof{$\times$}][c]{0} &  \times & \makebox[\widthof{$\times$}][c]{0} 
\\
 \makebox[\widthof{$\times$}][c]{0} &\makebox[\widthof{$\times$}][c]{0}&  \times
\end{pmatrix}
\hspace{2mm}
\Gamma_2:
\begin{pmatrix}
 \makebox[\widthof{$\times$}][c]{0}&  \times& \makebox[\widthof{$\times$}][c]{0}
\\
 \makebox[\widthof{$\times$}][c]{0}&   \makebox[\widthof{$\times$}][c]{0}&  \times
\\
\makebox[\widthof{$\times$}][c]{0} &  \makebox[\widthof{$\times$}][c]{0} & \makebox[\widthof{$\times$}][c]{0}
\end{pmatrix}
\hspace{2mm}
\\
\Delta_1&:
\begin{pmatrix}
 \times & \makebox[\widthof{$\times$}][c]{0} & \makebox[\widthof{$\times$}][c]{0}
\\
\makebox[\widthof{$\times$}][c]{0} &  \times & \makebox[\widthof{$\times$}][c]{0} 
\\
 \makebox[\widthof{$\times$}][c]{0} &\makebox[\widthof{$\times$}][c]{0}&  \times
\end{pmatrix}
\hspace{1,5mm}
\Delta_2:
\begin{pmatrix}
\makebox[\widthof{$\times$}][c]{0} &\makebox[\widthof{$\times$}][c]{0} & \makebox[\widthof{$\times$}][c]{0}
\\
\times &\makebox[\widthof{$\times$}][c]{0} & \makebox[\widthof{$\times$}][c]{0}
\\
\makebox[\widthof{$\times$}][c]{0} & \times  &\makebox[\widthof{$\times$}][c]{0}
\end{pmatrix}.
\end{split}
\end{align}

By rephasings of the left- and right-handed quarks, we can select, at most, real values for eight of the ten Yukawa couplings. One viable choice is to use real values for all couplings apart from the ones in $\Gamma_2$. With this, we have a reduction from 20 real parameters, down to 12, in the quark sector. For the lepton sector, there are, for all models in this paper, enough degrees of freedom to remove the complex phase for all couplings.

From Eq.~\eqref{eq:textureM2}, we see that there are FCNCs in all of the quark sectors, as neither of the sets $\{\Gamma_i\Gamma_j^\dagger\}$, $\{\Gamma_i^\dagger\Gamma_j\}$, $\{\Delta_i\Delta_j^\dagger\}$ and $\{\Delta_i^\dagger\Delta_j\}$ are Abelian (proof given in Ref.~\cite{Gatto:1979mr}). Nevertheless, the small number of free parameters allows us to extract some additional information. Using the principal invariants for $H_u=M_uM^\dagger_u$, we can reduce the number of free parameters to two. The relatively compact solutions for the up-quark sector read

\begin{align}
\begin{split}
\left|{M_u}\right|_{21}^2&=-a\pm\sqrt{b+a^2}
\\[0.2mm]
\left|M_u\right|_{32}^2&=-a+(M_u)_{11}^2-(M_u)_{33}^2\mp\sqrt{b+a^2}
\\[1mm]
(M_u)_{22}&= {m_u m_c m_t}/{(M_u)_{11}(M_u)_{33}}, 
\end{split}
\end{align}
with $a$ and $b$ defined, respectively, as

\begin{align}
\begin{split}
a&\equiv\frac{m_u^2m_c^2m_t^2}{2(M_u)_{11}^2(M_u)_{33}^2}-\frac{1}{2}\sum_{i=u,c,t} m_i^2+(M_u)_{11}^2,
\\
b&\equiv-\frac{1}{(M_u)_{11}^2} \prod_{i=u,c,t} \left(m_i^2-(M_u)_{11}^2\right). 
\end{split}
\end{align}
The same expression can be used for the $M_d$ mass matrix, with the replacement $(M_u)_{ii}\rightarrow (M_d)_{ii}$ and $|M_u|_{21,32}\rightarrow |M_d|_{12,23}$. In the down-quark sector, there are still the phases of the elements in $\Gamma_2$ that remain free.

The sources of FCNCs in the gauge sector are given by
\begin{align}
\begin{split}
\Xi_{\mathrm{{u_L}}}&=x\hspace{0.5mm}\mathbb{1}+ (y-x)\hspace{0.5mm}(P_2^{[\mathrm{{u_L}}]}+2P_3^{[\mathrm{{u_L}}]}),\\
\Xi_{\mathrm{{d_L}}}&=x\hspace{0.5mm}\mathbb{1}+ (y-x)\hspace{0.5mm}V^\dagger(P_2^{[\mathrm{{u_L}}]}+2P_3^{[\mathrm{{u_L}}]})V,\\
\Xi_{\mathrm{{u_{R}}}}&=(x+ 3y)\hspace{0.5mm}\mathbb{1}+ (y-x)\hspace{0.5mm}(P_2^{[\mathrm{{u_R}}]}+2P_3^{[\mathrm{{u_R}}]}),\\
\Xi_{\mathrm{{d_{R}}}}&=(x- 3y)\hspace{0.5mm}\mathbb{1}+ (y-x)\hspace{0.5mm}(P_2^{[\mathrm{{d_R}}]}+2P_3^{[\mathrm{{d_R}}]}),
\end{split}
\end{align}
with $x\neq y$ and with the projectors in the new basis defined as in Ref.~\cite{Alves:2018kjr}, namely,

\begin{align}
\label{eq:projectors}
\begin{split}
P_a^{[u_\mathrm{L}]}&\equiv U_{u_\mathrm{L}}^\dagger P_aU_{u_\mathrm{L}}, \;\;P_a^{[u_\mathrm{R}]}\equiv U_{u_\mathrm{R}}^\dagger P_aU_{u_\mathrm{R}},
\\
P_a^{[d_\mathrm{L}]}&\equiv V^\dagger P_a^{[u_\mathrm{L}]} V , \;\;\;P_a^{[d_\mathrm{R}]}\equiv U_{d_\mathrm{R}}^\dagger P_aU_{d_\mathrm{R}},
\end{split}
\end{align}
where $V$ is the CKM matrix {\small{$V=U_{\mathrm{uL}}^\dagger U_{\mathrm{dL}}$}} and with the projectors defined as $(P_a)_{ij}\equiv\delta_{ij}\delta_{ja}$, with no summation over repeated indices and with $i,j,a$ taking on values from one to three. For the scalar sector, the relevant sources of FCNCs can be expressed as

\begin{align}
\begin{split}
\frac{v_2}{\sqrt{2}}\mathcal{K}_u=&P_2^{[u_\mathrm{L}]}D_uP_1^{[u_\mathrm{R}]}+P_3^{[u_\mathrm{L}]}D_uP_2^{[u_\mathrm{R}]},\\
\frac{v_2}{\sqrt{2}}\mathcal{K}_d=&V^\dagger \big(P_1^{[u_\mathrm{L}]}VD_dP_2^{[u_\mathrm{R}]}+P_2^{[u_\mathrm{L}]}VD_dP_3^{[u_\mathrm{R}]}\big)\,.
\end{split}
\end{align}

\begin{figure}[ht!]
	\centering
		\includegraphics[width=0.52\textwidth]{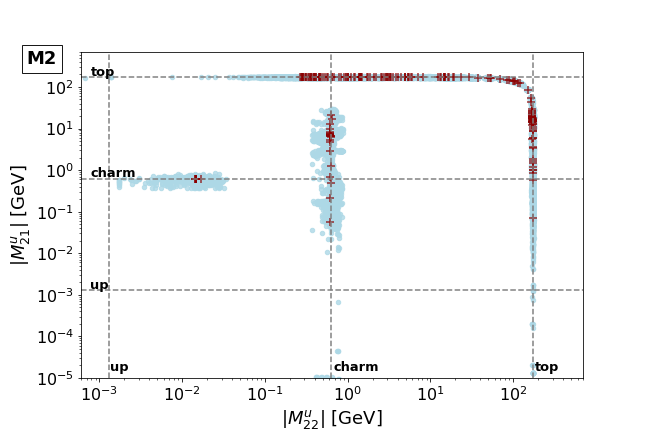}
	\caption{Preferred magnitudes for elements of the up quark mass matrix in the flavor basis for model M2. Red crosses correspond to regions with all deviations below $1\sigma$, and blue to deviations below $3\sigma$.} 
	\label{fig:ScatterM2MassesAndMixings}
\end{figure}

In Fig.~\ref{fig:ScatterM2MassesAndMixings}, we plot the preferred magnitudes for elements $(M_u)_{22}$ and $(M_u)_{21}$ when fitting MMs. The red parameter points correspond to a region of parameter space with all deviations below $1\sigma$, while blue points have a deviation below $3\sigma$. In the plot, we see a strong preference toward either $(M_u)_{22}$ or $(M_u)_{21}$ being of the order of the top mass. Similar features can be found in the down sector when plotting $(M_d)_{23}$ as a function of $(M_d)_{22}$. 

\subsubsection{Models M3a and M3b}

For model M3a, the textures are given by

\begin{align*}
\begin{split}
\Gamma_1&:
\begin{pmatrix}
 \times &  \times&   \makebox[\widthof{$\times$}][c]{0}
\\
 \times &  \times&   \makebox[\widthof{$\times$}][c]{0}
\\
 \makebox[\widthof{$\times$}][c]{0} &   \makebox[\widthof{$\times$}][c]{0}&  \times
\end{pmatrix}
\hspace{2mm}
\Gamma_2:
\begin{pmatrix}
 \makebox[\widthof{$\times$}][c]{0}& \makebox[\widthof{$\times$}][c]{0}& \makebox[\widthof{$\times$}][c]{0}
\\
 \makebox[\widthof{$\times$}][c]{0}& \makebox[\widthof{$\times$}][c]{0}& \makebox[\widthof{$\times$}][c]{0}
\\
 \times &  \times & \makebox[\widthof{$\times$}][c]{0}
\end{pmatrix}
\hspace{2mm}
\\
\Delta_1&:
\begin{pmatrix}
 \times &  \times&   \makebox[\widthof{$\times$}][c]{0}
\\
 \times &  \times&   \makebox[\widthof{$\times$}][c]{0}
\\
 \makebox[\widthof{$\times$}][c]{0} &   \makebox[\widthof{$\times$}][c]{0}&  \times
\end{pmatrix}
\hspace{1,5mm}
\Delta_2:
\begin{pmatrix}
 \makebox[\widthof{$\times$}][c]{0} &\makebox[\widthof{$\times$}][c]{0} & \times 
\\
 \makebox[\widthof{$\times$}][c]{0}& \makebox[\widthof{$\times$}][c]{0}&  \times 
\\
 \makebox[\widthof{$\times$}][c]{0}& \makebox[\widthof{$\times$}][c]{0}& \makebox[\widthof{$\times$}][c]{0}
\end{pmatrix}.
\end{split}
\end{align*} 

For this model, there are again FCNCs in all quark sectors, but this time around, there are further restrictions. With the charges of the first two generations being degenerate for all quarks, i.e.~$q_1=q_2$, $u_1=u_2$ and $d_1=d_2$, we have that

\begin{align}
\begin{split}
\Xi_{\mathrm{{u_L}}}&=x\hspace{0.5mm}\mathbb{1}+ (y-x)\hspace{0.5mm}P_3^{[\mathrm{{u_L}}]},\\
\Xi_{\mathrm{{d_L}}}&=x\hspace{0.5mm}\mathbb{1}+ (y-x)\hspace{0.5mm}V^\dagger P_3^{[\mathrm{{u_L}}]}V,\\
\Xi_{\mathrm{u_R}}&=(3x+y)\hspace{0.5mm}\mathbb{1}+(y-x)\hspace{0.5mm} P_3^{[u_\mathrm{R}]},\\
\Xi_{\mathrm{d_R}}&=-(x+y)\hspace{0.5mm}\mathbb{1} +(y-x)\hspace{0.5mm} P_3^{[d_\mathrm{R}]},
\end{split}
\end{align}
with $x\neq y$, while, for the scalar-mediated FCNCs,

\begin{align}
\begin{split}
\mathcal{K}_u&=\frac{\sqrt{2}}{v_2}\left(\mathbb{1}-P_3^{[u_\mathrm{L}]}\right)D_uP_3^{[u_\mathrm{R}]},
\\
\mathcal{K}_d&=\frac{\sqrt{2}}{v_2}V^\dagger P_3^{[u_\mathrm{L}]} VD_d\left(\mathbb{1}-P_3^{[d_\mathrm{R}]}\right).
\end{split}
\end{align}

Model M3b, on the other hand, has the same expression for $\Xi$ as model M3a, while the scalar-mediated FCNCs are now given by

\begin{align}
\begin{split}
\frac{v_2}{\sqrt{2}}\mathcal{K}_q=(\mathbb{1} - P_3^{[q_\mathrm{L}]})D_q(\mathbb{1} - P_3^{[q_\mathrm{R}]})+P_3^{[q_\mathrm{L}]}D_q P_3^{[q_\mathrm{R}]}
\end{split}
\end{align}
with $q=u,d$.

\begin{figure}[ht!]
	\centering
	\begin{subfigure}[b]{0.235\textwidth}
		\includegraphics[width=\textwidth]{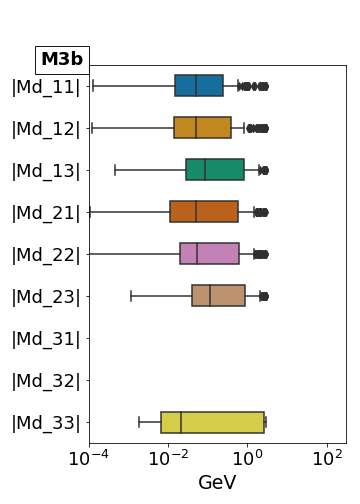}
	\end{subfigure}
	\begin{subfigure}[b]{0.235\textwidth}
		\includegraphics[width=\textwidth]{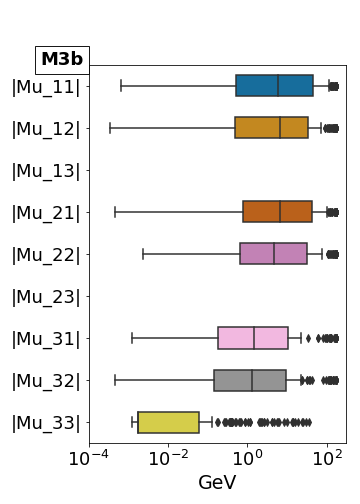}
	\end{subfigure}
	\caption{Preferred values for the elements of the up quark mass matrix (left) and down quark mass matrix (right) for model M3b. The included parameter points fit the MMs with a deviation below $1\sigma$.} 
	\label{fig:M3bBarPlot}
\end{figure}

Figure \ref{fig:M3bBarPlot} shows the preferred values for elements of the mass matrices. Here we see that, for model M3b, $(M_u)_{33}$ prefers having values around the up mass, while the remaining elements have a larger flexibility. The equivalent plot for model M3a shows a strong preference for $(M_d)_{33}$ to be of the order of the down quark mass. 

The figure is a so-called box plot, where the array of data is sorted by magnitude and then split into four equally sized parts, each referred to as a quartile. The black line within the box is the median of the dataset, while the box itself extends from the first to the third quartile. The so-called whiskers span from the smallest to the largest values in the set, out of the points not classified as outliers. An outlier is here defined as any point further away from the first or third quartile than one and a half times the length of the box. For a normal distribution, this would correspond to 0.7\% of the data. The outliers are marked as black dots in the figure and, due to the low number of good points in their vicinity, they indicate the fine-tuned regions of the parameter space.    

For both models M3a and M3b, at most eight out of the 14 Yukawa couplings can be set real from rephasings of the left- and right-handed quarks. There are, hence, 20 parameters in the quark sector, in addition to the three parameters in the lepton sector. One possible choice is for all couplings in $\Gamma_2$ and $\Delta_2$ to be real, in addition to the three diagonal couplings in $\Delta_1$ and $(\Gamma_1)_{33}$.




\subsection{Leptons that couple to both Higgs doublets}

This section contains all models fulfilling the constraints specified in Sec.~\ref{subsec:conditions}, while simultaneously having leptons that couple to both Higgs doublets. Contrary to the models discussed in the previous subsection, the inclusion of charged leptons carrying $U(1)^\prime$ charge is now crucial for all charge assignments.

\subsubsection{Models M4a and M4b}

For model M4a, the textures are given by

\begin{align*}
\begin{split}
\Gamma_1&:
\begin{pmatrix}
 \makebox[\widthof{$\times$}][c]{0} & \makebox[\widthof{$\times$}][c]{0} & \makebox[\widthof{$\times$}][c]{0}
\\
 \times & \times &  \makebox[\widthof{$\times$}][c]{0} 
\\
 \makebox[\widthof{$\times$}][c]{0}  & \makebox[\widthof{$\times$}][c]{0} & \times
\end{pmatrix}
\hspace{2mm}
\Gamma_2:
\begin{pmatrix}
\times& \times&  \makebox[\widthof{$\times$}][c]{0}
\\
 \makebox[\widthof{$\times$}][c]{0}&    \makebox[\widthof{$\times$}][c]{0}& \times
\\
\makebox[\widthof{$\times$}][c]{0} & \makebox[\widthof{$\times$}][c]{0} &  \makebox[\widthof{$\times$}][c]{0}
\end{pmatrix}
\hspace{2mm}
\\
\Delta_1&:
\begin{pmatrix}
\makebox[\widthof{$\times$}][c]{0} & \makebox[\widthof{$\times$}][c]{0} &\makebox[\widthof{$\times$}][c]{0} 
\\
\makebox[\widthof{$\times$}][c]{0} &\makebox[\widthof{$\times$}][c]{0} & \times
\\
\makebox[\widthof{$\times$}][c]{0} &\times &\makebox[\widthof{$\times$}][c]{0}
\end{pmatrix}
\hspace{1,5mm}
\Delta_2:
\begin{pmatrix}
\times &\makebox[\widthof{$\times$}][c]{0} & \makebox[\widthof{$\times$}][c]{0} 
\\
\makebox[\widthof{$\times$}][c]{0} & \makebox[\widthof{$\times$}][c]{0} & \makebox[\widthof{$\times$}][c]{0}
\\
\makebox[\widthof{$\times$}][c]{0} & \makebox[\widthof{$\times$}][c]{0}& \times
\end{pmatrix},
\end{split}
\end{align*}
with the corresponding charges specified in Table \ref{tab:charges}. Model M4a can have at most eight simultaneously real Yukawa couplings from rephasings of the quark fields. There are, hence, 12 parameters in the quark sector, in addition to the three parameters in the lepton sector. One allowed scenario is for all couplings in $\Gamma_1$, $\Delta_1$ and $\Delta_2$ to be real, in addition to $(\Gamma_2)_{11}$ (similar for M4b). 

\begin{figure}[ht!]
	\centering
		\includegraphics[width=0.52\textwidth]{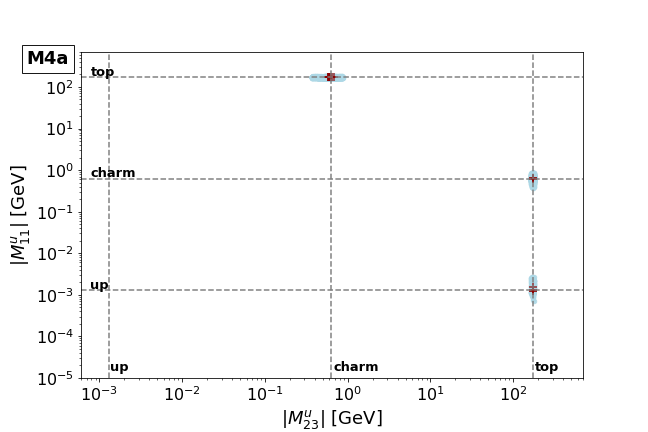}
	\caption{Preferred values for $(M_u)_{23}$ and $(M_u)_{11}$ for model M4a. Red crosses correspond to regions with all deviations below $1\sigma$, and blue ones to deviations below $3\sigma$.} 
	\label{fig:M4aScatterPlot}
\end{figure}

Since both combinations $M_uM_u^\dagger$ and $M_u^\dagger M_u $ are real and block diagonal, the diagonalization matrices $U_{\mathrm{uL}}$ and $U_{\mathrm{uR}}$ are simply given by a real orthogonal rotation. Using the same notation as in Eq.~\eqref{eq:orthogoRZ}, the rotation angle for $U_{\mathrm{uL}}$ will be given by

\begin{align}
\label{eq:mixingAngleM4a}
\begin{split}
\tan 2\theta_\mathrm{uL}=\frac{2\left(M_u\right)_{23}\left(M_u\right)_{33}}{\left(M_u\right)_{32}^2+\left(M_u\right)_{33}^2-\left(M_u\right)_{23}^2},
\end{split}
\end{align}
and the same for $\tan 2\theta_\mathrm{uR}$, but with $\left(M_u\right)_{23}$ and $\left(M_u\right)_{32}$ exchanged in all places. This can be further simplified by once again using the principal invariants, which has three possible solutions in the up sector, namely

\begin{align}
\begin{split}
(M_u)_{11}&=m_c, \;\;\;\;(M_u)_{32}=\frac{m_um_t}{(M_u)_{23}},
\\
(M_u)_{33}&=\frac{ \sqrt{ \Big((M_u)_{23}^2-m_u^2\Big)\Big(m_t^2-(M_u)_{23}^2\Big) } }{(M_u)_{23}},
\end{split}
\end{align}
and two equivalent expressions but with the masses exchanged as $m_c\leftrightarrow m_t$ and $m_c \leftrightarrow m_u$. Comparing this with the result of the scan in Fig.~\ref{fig:M4aScatterPlot}, we see, as expected, that $(M_u)_{11}$ always take on one of the three possible up-type quark masses. The gauge-mediated FCNC couplings are then given by

\begin{align}
\begin{split}
\Xi_{\mathrm{uL}}&=x\hspace{0.5mm}\mathbb{1}+(y-x)\hspace{0.5mm}\big(P_2^{[\mathrm{u_L}]}+2P_3^{[\mathrm{u_L}]}\big),\\
\Xi_{\mathrm{dL}}&=x\hspace{0.5mm}\mathbb{1}+(y-x)\hspace{0.5mm}V^\dagger\big(P_2^{[\mathrm{u_L}]}+2P_3^{[\mathrm{u_L}]}\big)V,\\
\Xi_{\mathrm{uR}}&=\frac{5x+7y}{3}\hspace{0.5mm}\mathbb{1}+(y-x)\hspace{0.5mm}\big(3P_2^{[\mathrm{u_R}]}+2P_3^{[\mathrm{u_R}]}\big),\\
\Xi_{\mathrm{dR}}&=\frac{x-7y}{3}\hspace{0.5mm}\mathbb{1}+(y-x)\hspace{0.5mm}P_3^{[\mathrm{d_R}]},
\end{split}
\end{align}
with $x\neq y$, and the scalar-mediated FCNCs by  

\begin{align}
\begin{split}
\mathcal{K}_u&=\frac{\sqrt{2}}{v_2}\left(P_1^{[u_\mathrm{L}]}D_uP_1^{[u_\mathrm{R}]}+P_3^{[u_\mathrm{L}]}D_uP_3^{[u_\mathrm{R}]}\right),\\
\mathcal{K}_d&=\frac{\sqrt{2}}{v_2}V^\dagger\left(P_1^{[u_\mathrm{L}]}VD_d+(P_2^{[u_\mathrm{L}]}-P_1^{[u_\mathrm{L}]})VD_dP_3^{[d_\mathrm{R}]}\right).
\end{split}
\end{align}
For model M4b, we have a similar scenario but with the roles of the up and down sectors exchanged.

\subsubsection{Model M5}

Next, we have a so-called "right model of type E" \cite{Alves:2018kjr}, with the textures

\begin{align*}
\begin{split}
\Gamma_1&:
\begin{pmatrix}
\times& \times & \makebox[\widthof{$\times$}][c]{0}
\\
\times &\times & \makebox[\widthof{$\times$}][c]{0}
\\
\times &\times&  \makebox[\widthof{$\times$}][c]{0}
\end{pmatrix}
\hspace{2mm}
\Gamma_2:
\begin{pmatrix}
 \makebox[\widthof{$\times$}][c]{0}&  \makebox[\widthof{$\times$}][c]{0}&  \times
\\
 \makebox[\widthof{$\times$}][c]{0}&    \makebox[\widthof{$\times$}][c]{0}& \times
\\
\makebox[\widthof{$\times$}][c]{0} & \makebox[\widthof{$\times$}][c]{0} & \times
\end{pmatrix}
\hspace{2mm}
\\
\Delta_1&:
\begin{pmatrix}
\times& \times & \makebox[\widthof{$\times$}][c]{0}
\\
\times &\times & \makebox[\widthof{$\times$}][c]{0}
\\
\times &\times&  \makebox[\widthof{$\times$}][c]{0}
\end{pmatrix}
\hspace{1,5mm}
\Delta_2:
\begin{pmatrix}
 \makebox[\widthof{$\times$}][c]{0}&  \makebox[\widthof{$\times$}][c]{0}&  \times
\\
 \makebox[\widthof{$\times$}][c]{0}&    \makebox[\widthof{$\times$}][c]{0}& \times
\\
\makebox[\widthof{$\times$}][c]{0} & \makebox[\widthof{$\times$}][c]{0} & \times
\end{pmatrix},
\end{split}
\end{align*}
and with no FCNCs in the left-handed quark sectors, as the charges of all left-handed quarks are degenerate. For the right-handed quarks, on the other hand, there are only two degenerate charges, such that $\Xi_\mathrm{uR}$ and $\Xi_\mathrm{dR}$ are of the form, respectively, 

\begin{align}
\begin{split}
\Xi_{\mathrm{uR}}&=y\hspace{0.5mm}\mathbb{1}-3(2x+y)\hspace{0.5mm}P_3^{[\mathrm{uR}]},\\
\Xi_{\mathrm{dR}}&=(2x-y)\hspace{0.5mm}\mathbb{1}+3(2x+y)\hspace{0.5mm}P_3^{[\mathrm{dR}]}.
\end{split}
\end{align} 
Similarly, for the scalar sector, we have

\begin{align}
\label{eq:NuNdnew}
\begin{split}
\mathcal{K}_u=\frac{\sqrt{2}}{v_2}D_u P_3^{[u_\mathrm{R}]}\quad\text{and}\quad \mathcal{K}_d=\frac{\sqrt{2}}{v_2}D_d P_3^{[d_\mathrm{R}]}.
\end{split}
\end{align}

As stated at the beginning of this section, for model M5 we use the physical quark masses and mixing matrix as inputs. Since the left-handed quark doublet charges are fully degenerate under $U(1)^\prime$, any weak basis transformation (WBT) in this sector will never affect any observable. Therefore, applying the WBT {{$q_\mathrm{L}^0\rightarrow U_\mathrm{uL}q_\mathrm{L}^0$}}, we are instead in the basis

\begin{align}
\begin{split}
M_u=D_uU_\mathrm{uR}, \;\;M_d=V^\dagger D_dU_\mathrm{dR} \,.
\end{split}
\end{align}
The only free quantities are the right-handed rotation matrices. The Yukawa couplings get, therefore, fully determined by the above mass matrices.  

\subsubsection{Models M6a and M6b}

Model M6a is the familiar BGL model \cite{Branco:1996bq},

\begin{align*}
\begin{split}
\Gamma_1&:
\begin{pmatrix}
\times& \times & \times
\\
\times &\times & \times
\\
 \makebox[\widthof{$\times$}][c]{0} & \makebox[\widthof{$\times$}][c]{0}&  \makebox[\widthof{$\times$}][c]{0}
\end{pmatrix}
\hspace{2mm}
\Gamma_2:
\begin{pmatrix}
 \makebox[\widthof{$\times$}][c]{0}&  \makebox[\widthof{$\times$}][c]{0}& \makebox[\widthof{$\times$}][c]{0}
\\
 \makebox[\widthof{$\times$}][c]{0}&    \makebox[\widthof{$\times$}][c]{0}& \makebox[\widthof{$\times$}][c]{0}
\\
\times & \times & \times
\end{pmatrix}
\hspace{2mm}
\\
\Delta_1&:
\begin{pmatrix}
\times& \times &\makebox[\widthof{$\times$}][c]{0}
\\
\times&\times & \makebox[\widthof{$\times$}][c]{0}
\\
\makebox[\widthof{$\times$}][c]{0} &\makebox[\widthof{$\times$}][c]{0} &\makebox[\widthof{$\times$}][c]{0}
\end{pmatrix}
\hspace{1,5mm}
\Delta_2:
\begin{pmatrix}
\makebox[\widthof{$\times$}][c]{0} &\makebox[\widthof{$\times$}][c]{0} & \makebox[\widthof{$\times$}][c]{0} 
\\
\makebox[\widthof{$\times$}][c]{0}&\makebox[\widthof{$\times$}][c]{0}  & \makebox[\widthof{$\times$}][c]{0}
\\
\makebox[\widthof{$\times$}][c]{0} & \makebox[\widthof{$\times$}][c]{0}& \times
\end{pmatrix},
\end{split}
\end{align*}
with the corresponding charges specified in Table \ref{tab:charges} and with FCNCs only in the left-handed down sector. 


As is well known for BGL-type models, $U_\mathrm{uL}$ and $U_\mathrm{uR}$ are block diagonal. In the right-handed down sector, all charges are degenerate, leading to a diagonal $\Xi_\mathrm{dR}$. For the other three quark sectors, on the other hand, there are two degenerate charges yielding 

\begin{align}
\begin{split}
\Xi_\mathrm{uL}&=x\hspace{0.5mm}\mathbb{1}+(y-x)\hspace{0.5mm}P_a,\\
\Xi_\mathrm{dL}&=x\hspace{0.5mm}\mathbb{1}+(y-x)\hspace{0.5mm}V^\dagger P_a V,\\
\Xi_\mathrm{uR}&=\frac{2}{3}(5x+y)\hspace{0.5mm}\mathbb{1}+2(y-x)\hspace{0.5mm}P_a,
\end{split}
\end{align}
with the projector index $a=1,2,3$ for when $(M_u)_{33}=m_u,m_c,m_t$, respectively. For the scalar-mediated FCNCs, we have

\begin{align}
\begin{split}
\mathcal{K}_u=\frac{\sqrt{2}}{v_2} P_a D_u \quad \text{and}\quad \mathcal{K}_d=\frac{\sqrt{2}}{v_2}V^\dagger P_a V D_d\,,
\end{split}
\end{align}
with $a$ defined as before.

Model M6b obeys similar relations, but with the role of the up and down sectors exchanged. Here the FCNCs are instead limited to the left-handed up sector.
For models M6a and M6b, there are, hence, no new parameters. 

Here, we can again choose a basis in which the expressions for the mass matrices are simplified. Besides the WBT used for model M5, we can now transform also the right-handed quarks such that in the new basis

\begin{align}
\begin{split}
M_u=D_u, \;\;M_d=V^\dagger D_d \,,
\end{split}
\end{align} 
without influencing the conserved currents, due to the way the degeneracy of charges coincides with the block diagonality of the right-handed mixing matrices.  

\subsubsection{Models M7a and M7b}

\begin{figure}[ht!]
	\centering
		\includegraphics[width=0.52\textwidth]{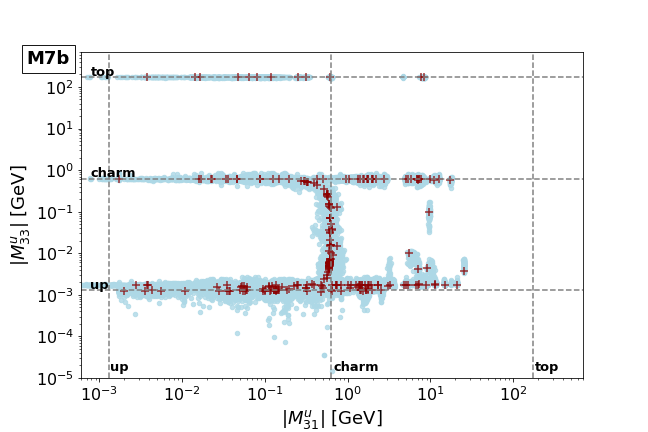}
	\caption{Preferred values for $(M_u)_{33}$ and $(M_u)_{31}$ in model M7b. Here, red crosses correspond to parameter points with all deviations smaller than $1\sigma$, and blue ones to all deviations below $3\sigma$.} 
	\label{fig:M7bScatterPlot}
\end{figure}

Finally, model M7a has the textures 

\begin{align*}
\begin{split}
\Gamma_1&:
\begin{pmatrix}
\times &\makebox[\widthof{$\times$}][c]{0} & \makebox[\widthof{$\times$}][c]{0} 
\\
\times&\makebox[\widthof{$\times$}][c]{0}  & \makebox[\widthof{$\times$}][c]{0}
\\
\makebox[\widthof{$\times$}][c]{0} & \makebox[\widthof{$\times$}][c]{0}& \times
\end{pmatrix}
\hspace{2mm}
\Gamma_2:
\begin{pmatrix}
\makebox[\widthof{$\times$}][c]{0}& \times &\makebox[\widthof{$\times$}][c]{0}
\\
\makebox[\widthof{$\times$}][c]{0}&\times & \makebox[\widthof{$\times$}][c]{0}
\\
\times &\makebox[\widthof{$\times$}][c]{0} &\makebox[\widthof{$\times$}][c]{0}
\end{pmatrix}
\hspace{2mm}
\\
\Delta_1&:
\begin{pmatrix}
 \makebox[\widthof{$\times$}][c]{0}&  \makebox[\widthof{$\times$}][c]{0}& \times
\\
 \makebox[\widthof{$\times$}][c]{0}&    \makebox[\widthof{$\times$}][c]{0}& \times
\\
\times & \times &  \makebox[\widthof{$\times$}][c]{0}
\end{pmatrix}
\hspace{1,5mm}
\Delta_2:
\begin{pmatrix}
\times& \times &  \makebox[\widthof{$\times$}][c]{0}
\\
\times &\times &  \makebox[\widthof{$\times$}][c]{0}
\\
 \makebox[\widthof{$\times$}][c]{0} & \makebox[\widthof{$\times$}][c]{0}&  \makebox[\widthof{$\times$}][c]{0}
\end{pmatrix},
\end{split}
\end{align*}
with the corresponding charges specified in Table \ref{tab:charges} and with FCNCs in all quark sectors. Here, there are two degenerate charges in all quark sectors except for the right-handed down quarks, such that the FCNC sources in the gauge sector are given by 

\begin{align}
\begin{split}
\Xi_\mathrm{uL}&=x\hspace{0.5mm}\mathbb{1}+(y-x)\hspace{0.5mm}P_3^{[\mathrm{u_L}]},
\\
\Xi_\mathrm{dL}&=x\hspace{0.5mm}\mathbb{1}+(y-x)\hspace{0.5mm}V^\dagger P_3^{[\mathrm{u_L}]} V,
\\
\Xi_\mathrm{uR}&=\frac{7x+5y}{3}\hspace{0.5mm}\mathbb{1}-(y-x)\hspace{0.5mm}P_3^{[\mathrm{u_R}]},\\
\Xi_\mathrm{dR}&=-\frac{4x+2y}{3}\hspace{0.5mm}\mathbb{1}+(y-x)\hspace{0.5mm}(P_3^{[\mathrm{u_R}]}-P_2^{[\mathrm{u_R}]}),
\end{split}
\end{align}
with $x\neq y$, while for the scalar-sector we get

\begin{align}
\begin{split}
\frac{v_2}{\sqrt{2}}\mathcal{K}_u&=(\mathbb{1}-P_3^{[u_\mathrm{L}]})D_u(\mathbb{1}-P_3^{[u_\mathrm{R}]}),\\
\frac{v_2}{\sqrt{2}}\mathcal{K}_d&=D_dP_2^{[d_\mathrm{R}]}-V^\dagger P_3^{[u_\mathrm{L}]}VD_d(P_1^{[d_\mathrm{R}]}-P_2^{[d_\mathrm{R}]}).
\end{split}
\end{align}
Model M7b obeys similar relations but with the down and up sector exchanged. In Fig.~\ref{fig:M7bScatterPlot}, we show the preferred values for entries of the up quark mass matrices, where red crosses correspond to regions with all deviations below $1\sigma$, and blue points to a deviation below $3\sigma$. Similar features can be found in the down sector of M7a.


For models M7a and M7b, there are, at most, nine out of 14 Yukawa couplings that can be set real from rephasings of the quark fields. As such, there are 19 parameters in the quark sector. One allowed selection of real entries correspond to all couplings in $\Gamma_2$ and $\Delta_2$, in addition to $(\Delta_1)_{12}$ and $(\Delta_1)_{31}$.




\subsection{Model tendencies}
The overall tendencies for fitting the quark MMs for the models are gathered in Table \ref{tab:MassesMixingFITs}.  Here, the second column displays the sum of all $\chi^2$ for the very best point for each model, where we see that all models contain regions in parameter space with an excellent agreement with the quarks MMs. The third column instead presents the percentage of points with an individual $\chi^2$ below one, from which we can tell that model M3a has the most preferable landscape, while model M4a has the least preferable one. And, finally, the fourth column presents the two most difficult observables to fit. For most models, this is the strange mass. Note that the largest individual deviation is still always below $5\sigma$.

\begin{table}
	\begin{tabular}{llll}
		\rule{0pt}{2.8ex}{\small{Model}}\;\;\;\;\; &  {\small{$\chi^2_{min}$}} \;\;\;\;\;\;\;\;\;\;\;  &  {\small{$\chi_i^2<1$}} \;\;\;\;\;  &  {\small{$1\leq\chi_i^2\leq 25$}} \;\;\;\;\;\;\; \\
		\hline\hline\\[-2mm]
		\rule{0pt}{2.8ex}{\small{{M2}}}  & $1.4\times 10^{-3}$  & $77\%$   & $\begin{array}{l}
		78\%\; [m_s]\\
		9\%\;\;\; [m_d]
		\end{array}$ \\[4mm]
		\rule{0pt}{2.8ex}{\small{{M3a}}}  & $1.8\times 10^{-3}$  & $93\%$   & $\begin{array}{l}
		64\%\; [m_b]\\
		23\%\; [m_s]
		\end{array}$ \\[4mm]
		\rule{0pt}{2.8ex}{\small{{M3b}}}  & $4.2\times 10^{-3}$  & $88\%$   & $\begin{array}{l}
		33\%\; [m_d]\\
		20\%\; [m_{s,b}]
		\end{array}$ \\[4mm]
		\rule{0pt}{2.8ex}{\small{{M4a}}}  & $3.8\times 10^{-3}$  & $60\%$   & $\begin{array}{l}
		84\%\; [m_s]\\
		11\%\; [|V_{td}|]
		\end{array}$ \\[4mm]
		\rule{0pt}{2.8ex}{\small{{M4b}}}  & $4.5\times 10^{-3}$  & $77\%$   & $\begin{array}{l}
		70\%\; [m_c]\\
		17\%\; [m_t]
		\end{array}$ \\[4mm]
		\rule{0pt}{2.8ex}{\small{{M7a}}}   & $2.9\times 10^{-3}$  & $88\%$   & $\begin{array}{l}
		89\%\;[m_s]\\
		11\%\;[m_c]
		\end{array}$ \\[4mm]
		\rule{0pt}{2.8ex}{\small{{M7b}}}   & $1.8\times 10^{-3}$  & $81\%$   & $\begin{array}{l}
		67\%\; [m_s]\\
		9\%\;\;\; [m_{t,u}]
		\end{array}$ \\[4mm]
		\hline
	\end{tabular}
	\caption{Display of the effort required for fitting MMs for the various models. The second column presents the total $\chi^2$ of the best-fit point, the third column shows the percentage of points with a $\chi^2$ below one, and the fourth column gives the observables with the most significant deviations from the data, for the points with a deviation in the range $1\sigma$-$5\sigma$.}
	\label{tab:MassesMixingFITs}
\end{table}

\section{Scan and phenomenological tests}\label{s:pheno}  

In this section, we present the details on the scan performed for each of the models found in Sec.~\ref{s:AnomalyFree}. For all models except M5 and M6, we have given as input the values of the best-fit points from the masses and mixing scan. We are then left to range over: 

\begin{itemize}
	\item neutral $Z^\prime$ gauge coupling $g^\prime \in [5\times 10^{-4},1]$;	
	\item scalar singlet VEV $v_S \in [10^2,10^6]$ GeV; 
	\item dimensionless scalar parameters $\lambda_{12,S,Si}\in [-1,1]$; and	
	\item dimensionful scalar parameter $a_1\in [-5,5]$ TeV.
\end{itemize}

The mass parameters $m_i^2$ and $m_S^2$ are fixed by the tadpole conditions, while $s_\theta$ and $f$ are left as free parameters (for more details, see Sec.~\ref{subsec:EW}). In the scan, we compute the masses for the neutral gauge bosons, neutral and charged scalars and their respective couplings to fermions and bosons. With that information, we then compute their corrections to a set of physical observables detailed below. 

For models M5 and M6, the approach is slightly different. As explained in the previous section, there is no need to perform a masses and mixing scan since these parameters can be given as input. In model M6, the Yukawa sector is fully determined by the known quark masses and CKM mixing, such that the only new free parameter is $\tan \beta$, for which we choose $\beta$ to be in the interval $[10^{-3},1.56]$. In model M5, we can once more set the quarks' masses and mixing to the physical ones, but we are still left with two generic unitary matrices in the up and down right-handed quark sectors. These unitary matrices are parameterized as $V_R=KR_{12}R_{13}R_{23}$, with $K$ a diagonal matrix with three phases bounded by $[10^{-1},2\pi[$ and $R_{ij}$ a complex rotation in the $ij$ plane with both the angle and phase constrained to the interval $[10^{-5},2\pi[$.

Next, we give a more detailed description of the observables used in the scan.

\subsection{Electroweak and low-energy constraints}\label{subsec:EW}

As already discussed in Sec.~\ref{s:U12HDM}, the presence of mixing between the "would-be" SM $\hat{Z}$ and the new gauge boson $\hat{Z}^\prime$ will, in turn, lead to modifications on the currents of the physical $Z$ boson. The electroweak sector of the SM has been probed with a very high level of accuracy, such that only a very small mixing is, in general, allowed. In our study of the EW constraints, we look at the $Z$-pole pseudo-observables
\begin{align*}
\begin{split}
\Gamma_Z^\mathrm{tot}:&\quad \text{Z total decay width}\,,\\
\sigma_\mathrm{had}:&\quad e^+e^- \text{cross section to hadrons} \,,\\
A_f:&\quad\text{parity-violating assym. } f=\{b,c,s,l\}\,,\\
A_f^\mathrm{FB}:&\quad \text{forward-backward assym. }f=\{b,c,s,l\}\,,\\
R_f:&\quad \text{partial width ratios }f=\{b,c,l\}\,.
\end{split}
\end{align*}
To compute these observables, we follow closely the \textsc{Leptop} computer program~\cite{Novikov:1995ap,Novikov:1995vu} approach, which includes one-loop EW corrections. In this approach $\bar{\alpha}$, $m_Z$ and $G_{F}$ are given as inputs, where $G_{F}$ is the Fermi coupling constant. However, since the presence of the new $Z^\prime$ boson alters the $Z$ mass, we include $m_Z$ and $\bar{\alpha}$ in the fit.
We then take 

\begin{align}
\label{eq:sinthetaDEF}
s_\theta^2 c_\theta^2=\frac{\pi \bar{\alpha}}{\sqrt{2}G_{F} m_\mathrm{Z}^2}
\end{align}
as the defining relation for $s_\theta$, where $\bar{\alpha}$ includes the running due to the three lepton and five "light" quark loops leading to $\bar{\alpha}^{-1}\equiv \alpha(m_Z^2)^{-1}=128.878\pm 0.090$~\cite{Novikov:1995vu}. The analytical definitions for these pseudo-observables are the standard ones, and their one-loop corrections can be found in Refs.~\cite{Novikov:1995ap,Novikov:1995vu}. The effects of the scalars were taken into account through the computation of the oblique parameters~\cite{Grimus:2008nb}, along the same lines as Burgess~\cite{Burgess:1993vc}.

We also compute the off-pole cross sections with both $Z$ and $Z^\prime$, for the final states $\mu^+\mu^-$, $\tau^+\tau^-$ and $q\bar{q}$, within the energy range $\sqrt{s}=[130,207]$ GeV given by LEP-II~\cite{Schael:2013ita}.

The previous observables do not really constrain the interactions of the $Z$ boson with the top quark. We have therefore included also the top width in the fit, in addition to the rare top decays $t\rightarrow Z q$ and $t\rightarrow h q$.

Low-energy constraints such as atomic parity violation, electric dipole moments (EDMs), and the muon magnetic moment can place significant bounds on the parameter space of our scenarios. Atomic parity violation for ${}^{133}_{\hspace{0.15cm} 55}\mathrm{Cs}$ and ${}^{205}_{\hspace{0.15cm} 81}\mathrm{Tl}$ have been included in the fit~\cite{Erler:2003yk,Ginges:2003qt}. For the neutron EDM, we have considered one-loop corrections from neutral and charged scalars~\cite{Jung:2013hka} and Barr-Zee diagrams~\cite{Barr:1990vd,Abe:2013qla}, while for the muon magnetic moment both the gauge and scalar one-loop contributions are taken into account~\cite{Jegerlehner:2009ry}. The EDM of the electron, on the other hand, is virtually zero since there are no FCNCs in the lepton sector.

\subsection{Meson constraints}\label{subsec:meson}

A flavor-changing mediator, scalar or vector, is strongly constrained from meson observables. In the fit we have included:

\begin{align*}
\begin{split}
\Delta M_{M^0}:&\quad \text{mass splitting~\cite{Buras:2012jb,Buras:2013rqa,Buras:2014zga,Golowich:2007ka,Golowich:2009ii}}\,,\\
\epsilon_K:&\quad \text{Kaon sector $CP$ asymmetry~\cite{Buras:2012jb,Buras:2013rqa,Buras:2014zga}}\,,\\
S_{\psi K_s}\,,\, S_{\psi\phi}:&\quad \text{$B$-sector $CP$-violating obs.~\cite{Buras:2012jb,Buras:2013rqa,Buras:2014zga}}\,,\\
B_{s}\rightarrow \mu\mu:&\quad \text{leptonic decay~\cite{Buras:2012ru,Buras:2013uqa,DeBruyn:2012wj,Aaij:2017vad}}\,,\\
B\rightarrow X_s\gamma:&\quad \text{radiative decay~\cite{Misiak:2015xwa,Buras:2011zb}}\,.
\end{split}
\end{align*}
The most dominant bounds usually come from the $\Delta F=2$ transitions observables, $\Delta M_{M^0}$ and $\epsilon_K$. The SM contribution for the $K^0$ and $B^0_q$ systems is dominated by the top-loop box diagrams, while the SM prediction for the $D^0$ system is less trivial. The short-distance effects are virtually zero, but long-distance contributions can be sizable (at the level of the current experimental values). However, long-distance effects are plagued with significant hadronic uncertainties, making it difficult to estimate the SM contribution. For this, we follow the approach of Refs.~\cite{Golowich:2007ka,Golowich:2009ii} and attribute the experimental values to short-distance NP. This will serve as an upper bound, such that NP contributions do not greatly exceed the current experimental bound.  Box diagrams for the charged Higgs were also included in the numerical scan; see~\cite{Crivellin:2013wna} for the general expressions. The hadronic matrix elements for $\Delta F=2$ processes are very important and were included together with the QCD running of the Wilson coefficients from the NP scale down to the meson scale; this was done in line with Refs.~\cite{Buras:2012jb,Buras:2013rqa,Buras:2014zga}. We took a conservative approach for the meson oscillation observables and for the Kaon sector, with the error for $\Delta M_K$ taken as $20\%$ of the central value~\cite{Tanabashi:2018oca}, while for $|\epsilon_K|$ we take the error to be $10\%$. We also use a $20\%$ error for the $D$-meson sector. For the $B$-meson sector, since it is less sensitive to long-distance contributions, we took the error for the $\Delta M_{B_q}$ to be on the $5\%$ level. 

In the scan, we have also included other meson observables that currently act as upper bounds, such as $K_L\rightarrow \mu^+\mu^-$~\cite{Ecker:1991ru,Isidori:2003ts,DAmbrosio:2017klp,Aaij:2017tia}, $K^{+}\rightarrow \pi^+\nu\bar{\nu}$ and $K_L\rightarrow \pi^0\nu\bar{\nu}$~\cite{Artamonov:2008qb,Bobeth:2016llm,Buras:2015qea}, $K_L\rightarrow \pi^0 e^+ e^-$ and $K_L\rightarrow \pi^0 \mu^+ \mu^-$~\cite{Mescia:2006jd,Blanke:2006eb,Buchalla:2003sj,AlaviHarati:2003mr,AlaviHarati:2000hs}, and $B_d\rightarrow \ell^+\ell^-$ and $B_s\rightarrow e^+e^-,\tau^+\tau^-$~\cite{Crivellin:2013wna,Tanabashi:2018oca}.

The $b\rightarrow s \ell\ell$ transitions are very relevant for these classes of models and have gotten the attention of a large community since the 2014 LHCb results~\cite{Aaij:2014ora} seemed to indicate a breaking of lepton universality. There have been several global fit analyses~\cite{Descotes-Genon:2013wba,Altmannshofer:2013foa,Altmannshofer:2014rta,Alguero:2019ptt}, pointing the preferred region for some of the four-Fermi effective operators. The experimental status is very likely to change in the next couple of years. Currently LHCb reports~\cite{Aaij:2019wad}

\begin{align}
\begin{split}
\frac{\mathrm{Br}(B_s\rightarrow K\mu^+\mu^-)}{\mathrm{Br}(B_s\rightarrow K e^+e^-)}\equiv R^{[1.1,6]}_K = 0.846^{+0.060+0.016}_{-0.054-0.014}\,,
\end{split}
\end{align}
suggesting a $2\sigma$-$3\sigma$ deviation from the SM, given $\left.R_K\right|_\mathrm{SM}\simeq 1+\mathcal{O}(m_\mu^2/m_b^2)$~\cite{Hiller:2003js}. The Belle Collaboration also measured this observable and gives values more compatible with the SM~\cite{Belle:conferenceNr2},  i.e.
\begin{align}
\begin{split}
R_K^{[1,6]} &= 0.98^{+0.27}_{-0.23}\pm 0.06\,,\\
R_K^{[q^2>14.18]} &= 1.11^{+0.29}_{-0.26}\pm 0.07\,.
\end{split}
\end{align}
In addition to these results, the Belle Collaboration has also reported the ratio for the $B_s\rightarrow K^\ast \ell\ell$ decays~\cite{Abdesselam:2019wac}: 
\begin{align}
\begin{split}
R_{K^\ast}^{[0.045,1.1]} &= 0.52^{+0.36}_{-0.26}\pm 0.05\,,\\
R_{K^\ast}^{[1.1,6]} &= 0.96^{+0.45}_{-0.29}\pm 0.11\,,\\
R_{K^\ast}^{[15,19]} &= 1.18^{+0.52}_{-0.32}\pm 0.10\,.
\end{split}
\end{align}
To compute the contributions of NP to these observables, we start by defining the effective Hamiltonian

\begin{align}
\label{eq:Heff_bsll}
\mathcal{H}_\mathrm{eff}^{b\rightarrow s\ell\ell}=-\frac{G_F}{\sqrt{2}}\frac{\alpha}{\pi}V_{tb}V_{ts}^\ast
\sum_i\left(C_i\mathcal{Q}_i+C_i^\prime\mathcal{Q}_i^\prime\right)\,,
\end{align}
with

\begin{align}
\begin{split}
\mathcal{Q}_9^{(\prime)}&=(\bar{s}\gamma_\mu P_{L(R)}b)(\bar{\ell}\gamma^\mu\ell)
\,,\\
\mathcal{Q}_{10}^{(\prime)}&=(\bar{s}\gamma_\mu P_{L(R)}b)(\bar{\ell}\gamma^\mu\gamma_5\ell)\,.
\end{split}
\end{align}
For the SM Wilson coefficients, we have $C_9^\mathrm{SM}\simeq -C_{10}^\mathrm{SM}\simeq 4.2$ and zero for their primed counterparts. The $R_K$ and $R_{K^\ast}$ ratios are then given by~\cite{Hiller:2014ula}

\begin{align}
\begin{split}
R_{K/K^\ast}&\simeq 1+\Delta_{\pm} + \Sigma_{\pm}\,,
\end{split}
\end{align}

with $\Delta_{\pm}$ the SM-NP interference term and $\Sigma_{\pm}$ the pure NP contribution. Their explicit form is given by

\begin{align}
\begin{split}
\Delta_\pm&\simeq 0.24\,
\mathrm{Re}\left\{
C_{9-10}^\mathrm{\mu\mu,NP} \pm C_{9-10}^\mathrm{\prime\mu\mu,NP}
\right\}
-(\mu\rightarrow e)
\end{split}
\end{align}
and

\begin{align}
\begin{split}
\Sigma_\pm&\simeq0.028\, \sum_{i=9,10}\left|C_i^\mathrm{\mu\mu,NP}\pm C_i^\mathrm{\prime\mu\mu,NP}\right|^2 -(\mu\rightarrow e)\,,
\end{split}
\end{align}
where we defined $C^{(\prime)}_{9-10}\equiv C^{(\prime)}_9-C^{(\prime)}_{10}$. These Wilson coefficients, in our framework, take the form

\begin{align}
\begin{split}
C^{(\prime)ij,NP}_{9}=-\frac{v^2\pi}{\alpha \lambda_t^{B_s}}\left[\frac{(Q^{d_{L(R)}}_{Z^0})_{sb}(Q^{e_{R}}_{Z^0}+ Q^{e_{L}}_{Z^0})_{ij}}{m_{Z^0}^2}\right]\,,
\end{split}
\end{align}
with $\lambda_t^{B_s}\equiv V_{tb}V_{ts}^\ast$ and $Z^0=\{Z,Z^\prime\}$, and equivalent for $C_{10}$, but with the sign in front $Q^{e_{L}}_{Z^0}$ flipped. The corresponding results are discussed in Sec.~\ref{subsec:flavorAno}.

The rare decay modes $B\rightarrow\{K,K^*,X_s\}\nu\bar{\nu}$ can be important tests for NP. The SM short-distance predictions were computed in Ref.~\cite{Altmannshofer:2009ma} and are still an order of magnitude below the experimental bounds~\cite{Barate:2000rc,Chen:2007zk,Aubert:2008am}, leaving plenty of room for NP. We can parameterize the NP effects through the ratios with the SM predictions; see Ref.~\cite{Altmannshofer:2009ma}.

\subsection{Collider constraints}

At the LHC, several direct searches for new neutral gauge and scalar bosons have been performed. As we shall see in the next section, due to the necessity of a large VEV for the scalar singlet $S$, the new scalar fields tend to be heavier than the $Z^\prime$ and, in most of the models, out of reach for the current colliders. We have therefore looked at only the collider constraints coming from direct searches of the $Z^\prime$ gauge boson. 

We follow a simplified approach by working in the narrow width approximation (NWA). For all models, this turns out to be a good approximation in the low $m_{Z^\prime}$ range, i.e.~below a few tens of TeV. Under the NWA we split the cross section under production and decay, i.e.

\begin{align}
\begin{split}
\left.\sigma(pp\rightarrow Z^\prime \rightarrow  \mathrm{X})\right|_\mathrm{NWA}=\sigma(pp\rightarrow Z^\prime)\mathrm{Br}(Z^\prime \rightarrow  \mathrm{X})\,,
\end{split}
\end{align}  
where the production cross section is given by~\cite{Carena:2004xs,Accomando:2010fz}

\begin{align}
\sigma(pp\rightarrow Z^\prime)=\frac{4}{3}\frac{\pi^2}{s}\sum_{q}\frac{\Gamma(Z^\prime\rightarrow q\bar{q})}{m_{Z^\prime}}\,\omega_q(s,m^2_{Z^\prime})\,,
\end{align}
with $q=u,c,d,s,b$. The integration over the parton distribution functions is included in the definition of $\omega_q(s,m^2_{Z^\prime})$ given in Ref.~\cite{Paz:2017tkr}; these were computed using the parton distribution function set \textsf{NNPDF23\_nnlo\_as\_0119\_qed}~\cite{Ball:2013hta}.

The experimental studies used in this analysis are presented in Tab.~\ref{tab:collider}.

\begin{table}
	\begin{tabular}{r|cl}
		\;\;\;\;\;\;\;Channel\;\;&$m_{Z^\prime}$ [GeV]&\;\;\;\;Collaboration\;\;\;\;\;\;\;\;\\
		\hline\hline\\[-2mm]
		$e^+e^-,\,\mu^+\mu^-$&[250,6000]&\;\;\;\; ATLAS@13TeV~\cite{Aad:2019fac}\\[1mm]
		&[150,3500]&\;\;\;\; ATLAS@8TeV~\cite{Aad:2014cka}\\[1mm]
		&[190,3000]&\;\;\;\; ATLAS@7TeV~\cite{Aad:2012hf}\\[2mm]
		$\tau^+\tau^-$&[200,4000]& \;\;\;\; ATLAS@13TeV~\cite{Aaboud:2017sjh}\\[1mm]
		&[500,2500]&\;\;\;\; ATLAS@8TeV~\cite{Aad:2015osa}\\[2mm]
		$t\bar{t}$\;\;&[500,5000]&\;\;\;\; ATLAS@13TeV~\cite{Aaboud:2019roo}\\[1mm]
		&[390,5000]&\;\;\;\; ATLAS@13TeV~\cite{Aaboud:2018mjh}\\[2mm]
		$jj$\;\;&[570,8100]&\;\;\;\; CMS@13TeV~\cite{CMS:2017xrr}\\[2mm]
		$W^+W^-$&[1190,4100]&\;\;\;\; CMS@13TeV~\cite{Sirunyan:2017acf}\\[1mm]
		&[550,4000]&\;\;\;\; CMS@13TeV~\cite{Sirunyan:2016cao}\\[1mm]
		&[300,5000]&\;\;\;\; ATLAS@13TeV~\cite{Aaboud:2017fgj}\\[2mm]
		$Zh\rightarrow Z b\bar{b}\, (c\bar{c})$&[500,5000]&\;\;\;\; ATLAS@13TeV~\cite{ATLAS:2017nxi}\\[2mm]
		$Z\gamma\rightarrow ee(\mu\mu)\gamma$&[350,4000]&\;\;\;\; CMS@13TeV~\cite{Sirunyan:2017hsb}\\[2mm]
		\hline\hline
	\end{tabular}
	\caption{LHC direct $Z^\prime$ searches used for the collider constraints in our scans. The first column shows the final state. The second and third columns give the covered $Z^\prime$ mass range and reference, respectively.}
	\label{tab:collider}
\end{table}

\section{Model constraints and predictions}\label{s:modelPred}

In this section, we study all models except for M1, as it has no FCNCs and has already been extensively studied elsewhere. For the numerical scans, we needed to fix the free charges of Table \ref{tab:charges}. The choice made was based on two simple requirements:

\begin{itemize}
\item[(a)] No $Z^\prime$ coupling to electrons, for both left- and right-handed chiralities. This choice helps to avoid the bounds coming from LEP.
\item[(b)] The largest charge in modulus is 1.
\end{itemize}
Under (a) and (b), we got the following assignments 

\begin{align}
\label{eq:chargeAssignments}
\begin{split}
  &\big(x,y,z\big)_\mathrm{M2}=\frac{1}{3}\big(1,0,3\big),\hspace{2mm} \big(x,y\big)_\mathrm{M5}=\frac{1}{12}\big(1,1\big),
  \\[1.5mm]
  &\big(x,y\big)_\mathrm{M4a}=\frac{1}{39}\big(10,1\big),\hspace{2mm} \big(x,y\big)_\mathrm{M4b}=\frac{1}{33}\big(-8,1\big), 
   \\[1.5mm]
   &\big(x,y\big)_\mathrm{M6a}=\frac{1}{30}\big(-2,7\big),\hspace{2mm} \big(x,y\big)_\mathrm{M6b}=\frac{1}{24}\big(-4,5\big), 
  \\[1.5mm]
 &\big(x,y\big)_\mathrm{M7a}=\frac{1}{15}\big(2,-7\big),\hspace{2mm}  \big(x,y\big)_\mathrm{M7b}=\frac{1}{21}\big(4,-5\big),
   \\[1.5mm]
    &\big(x,y,z\big)_\mathrm{M3ab}=\frac{1}{3}\big(1,-2,3\big),
\end{split}
\end{align}
with the corresponding charges given in Table \ref{tab:charges}.

\begin{figure}[ht!]
    \centering
        \includegraphics[width=0.42\textwidth]{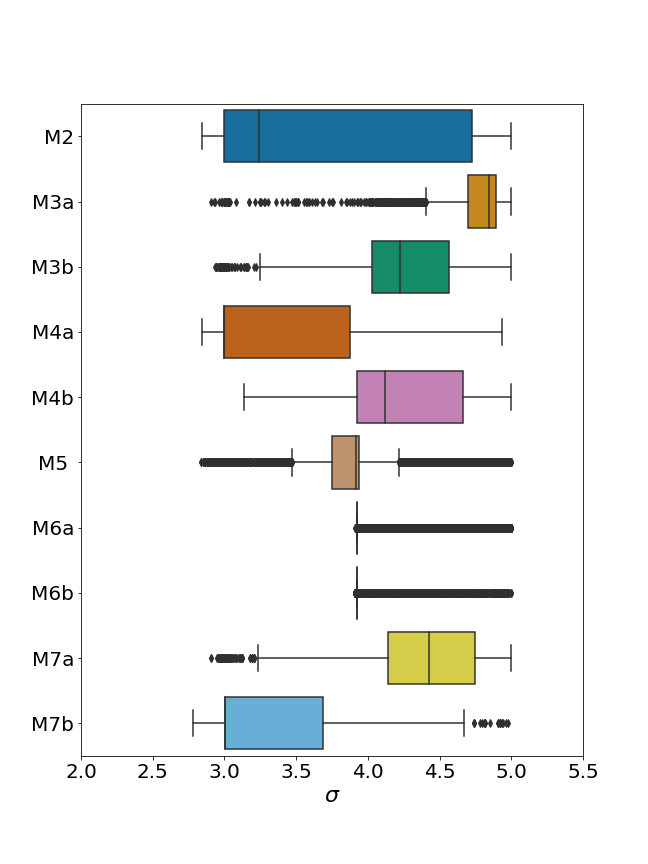}
    \caption{Figure displaying the ease for the various models in fitting all observational constraints, with the deviation in units of sigma on the $x$ axis.} 
    \label{fig:BarPlotSummary}
\end{figure}

Some general results for these charges are gathered in Fig.~\ref{fig:BarPlotSummary}, Table \ref{tab:FIT}, and Table \ref{tab:ScalarMasses}. Figure \ref{fig:BarPlotSummary} shows the ability of the various models to fit all observables. Here we see that, for the explored regions of parameter space, models M2, M3a, M3b, M4a, M5, M7a and M7b can all reach low deviations with data. In particular, models M2, M4a and M7b do so without the corresponding parameter points being outliers of the distributions. Surprisingly enough, the BGL model lacks parameter points below $\sim 3.9 \sigma$. This could be understood from the fact that, while the FCNCs are naturally suppressed in the BGL model, they are also fixed, which reduces the flexibility when fitting parameters. 

Table \ref{tab:FIT} displays again the effort required for fitting the various models. Here, we show points with an individual deviation below $5\sigma$ and $m_{Z^\prime}<30$ TeV. The second column contains the number of parameter points, out of the initial 300, that survived this requirement when fitting the MMs. The MM points were then used as input to the minimization of the remaining physical observables, with the number of surviving parameter points shown in the third column. Many of these points share the same MM input but differ in values for the remaining free parameters of the model. Next, the fourth column displays the most constraining collider channels, out of the ones in Table \ref{tab:collider}. The corresponding channel is displayed only whenever it actually excludes any of the parameter points from the scan, for that corresponding model. Overall, collider constraints are dominated by the $\mu^+\mu^-$ channel. We recall that for all model implementations chosen, the electron carries no charge under $U(1)^\prime$. Finally, the fifth column identifies the observables with the largest individual $\chi^2$, with the percentages reflecting their relative dominance. For example, for model M4b, $78\%$ of the FIT points had their largest deviation coming from the meson observable $\Delta M_K$. 


While a more complete scanning of the parameters space would be desirable, the results in Table \ref{tab:FIT} already show some of the model tendencies. For almost all models, $\Delta M_K$ is the observable with the highest deviation. However, since the meson sector is plagued with uncertainties on hadronic matrix elements and long-distance contributions, models with deviations around $4\sigma$ should not be completely discarded. Such sizable deviations in the EW sector are more difficult to accommodate making, for example, model M4a less appealing.

\begin{table}
	\begin{tabular}{lllll}
		&  {\small{MM }} &  {\small{FIT}}& {\small{Collider}} &  \\[-2mm]
		\rule{0pt}{2.8ex}{\small{Model}}\;\;\;\;\; &  {\small{points}} \;\;\;\;\; &  {\small{points}} \;\;\;\;\;  &  {\small{channel}} \;\;\;\; &  {\small{max($\chi^2_i$)}} \;\;\;\;\;\;\;\\
		\hline\hline\\[-2mm]
		M2&264&1.2$\times10^3$&--&
		$\begin{array}{ll}
		64\%&[\Delta M_K]\\
		27\%&[\sigma_\mathrm{had}]\\
		7\%&[\epsilon_K]
		\end{array}$\\[6mm]
		M3a&291&8.1$\times10^3$&$\mu^+\mu^-$&
		$\begin{array}{ll}
		98\%&[\Delta M_K]\\
		1\%&[\Delta M_D]
		\end{array}$\\[4mm]
		M3b&237&4.7$\times10^3$&$
		\begin{array}{l}
		\mu^+\mu^-\\
		\tau^+\tau^-
		\end{array}
		$&
		$\begin{array}{ll}
		91\%&[\Delta M_K]\\
		4\%&[\epsilon_K]\\
		4\%&[\sigma_\mathrm{had}]
		\end{array}$\\[6mm]
		M4a&231&1.4$\times10^3$&$
		\begin{array}{l}
		\mu^+\mu^-\\
		\tau^+\tau^-
		\end{array}
		$&
		$\begin{array}{ll}
		60\%&[\sigma_\mathrm{had}]\\
		30\%&[m_s]\\
		5\%&[\epsilon_K]
		\end{array}$\\[6mm]
		M4b&215&1.8$\times10^3$&$
		\mu^+\mu^-
		$&
		$\begin{array}{ll}
		78\%&[\Delta M_K]\\
		13\%&[m_c]\\
		3\%&[\Delta M_D]
		\end{array}$\\[6mm]
		M5&--&9.6$\times10^3$&$
		\begin{array}{l}
		\mu^+\mu^-\\
		\tau^+\tau^-
		\end{array}
		$&
		$\begin{array}{ll}
		56\%&[\Delta M_K]\\
		15\%&[m_Z]\\
		7\%&[\sigma_\mathrm{had}]
		\end{array}$\\[6mm]
		M6a&--&63$\times10^3$&$
		\begin{array}{l}
		\mu^+\mu^-\\
		\tau^+\tau^-\\
		Zh
		\end{array}
		$&
		$\begin{array}{ll}
		82\%&[\Delta M_K]\\
		12\%&[m_Z]\\
		3\%&[\Gamma_Z]
		\end{array}$\\[6mm]
		M6b&--&16$\times10^3$&$
		\begin{array}{l}
		\mu^+\mu^-\\
		\tau^+\tau^-\\
		Zh
		\end{array}
		$&
		$\begin{array}{ll}
		92\%&[\Delta M_K]\\
		5\%&[m_Z]\\
		2\%&[\Gamma_Z]
		\end{array}$\\[6mm]
		M7a&293&7.7$\times10^3$&$
		\mu^+\mu^-
		$&
		$\begin{array}{ll}
		88\%&[\Delta M_K]\\
		5\%&[\epsilon_K]
		\end{array}$\\[4mm]
		M7b&289&6.7$\times10^2$&--&
		$\begin{array}{ll}
		50\%&[\Delta M_K]\\
		36\%&[\sigma_\mathrm{had}]\\
		6\%&[\epsilon_K]
		\end{array}$\\
		\hline
	\end{tabular}
	\caption{Table showing the effort required to fit the various models, with $\chi_i^2<25$ and $m_{Z^\prime}<30$ TeV. 
	}
	\label{tab:FIT}
\end{table} 

In general, all the models considered require a vast hierarchy between the EW scale and the VEV of the scalar singlet, resulting in the decoupling of the neutral scalar singlet ($H_3^0$) and a full doublet of scalars ($H_1^0$, $H_2^0$ and $H^+$). In Table \ref{tab:ScalarMasses}, we have presented the lowest possible scalar masses obtained in the scan for the various models. Here, we see that, even for the parameter points with the lowest scalar masses, the scale hierarchies are evident. Only models M3b, M5, M6a, M6b and M7a have scalar masses below 4 TeV. These are, however, quite isolated points, as for all models except M6, this amounts to less than $1\%$ of the FIT points in Table \ref{tab:FIT}. For models M6a and M6b, however, $v_S$ can be of the order of a few TeV, allowing the scalar masses to come down to a few hundred GeV. For M6a these amount to around $3\%$ of the FIT points and $10\%$ for model M6b. 

As already mentioned, we have included only the direct searches on $Z^\prime$, such that similar searches for scalars could easily rule out these low-mass points. Therefore, Table \ref{tab:ScalarMasses} should serve only as an indication for which models we might expect relatively light scalars. Also, the results presented here correspond to the charge assignments given in Eq.~\eqref{eq:chargeAssignments}. A different choice could help lower the bound for the scalar masses. For example, in model M5, there were many allowed points with scalar masses in the range $[500,10^3]$ GeV. All these were excluded only once the collider searches were included. A different coupling to light quarks could then easily have revived these points.

\begin{table}
	\begin{tabular}{lllll}
		&  \multicolumn{4}{c}{Lowest Scalar Masses [TeV]} \\
		\cline{2-5}
		\rule{0pt}{2.8ex}{\small{Model}}\;\;\;\;\;\;\;\;\;\;\;\; &  {\small{$H^0_1$}} \;\;\;\;\;\;\;\;\;\;\;\;  &  {\small{$H^0_2$}} \;\;\;\;\;\;\;\;\;\;\;\;  &  {\small{$H^0_3$}} \;\;\;\;\;\;\;\;\;\;\;\;  &  {\small{$H^+$}}\\
		\hline\hline
		\rule{0pt}{2.8ex}{\small{{M2}}}  & 11  & 11   & 42  & 11   \\
		\rule{0pt}{2.8ex}{\small{{M3a}}}  & 4.3  & 4.9   & 21  & 4.9   \\
		\rule{0pt}{2.8ex}{\small{{M3b}}}  & 1.3  & 2.9   & 7.8  & 2.9   \\
		\rule{0pt}{2.8ex}{\small{{M4a}}}  & 14  & 14   & 41  & 14   \\
		\rule{0pt}{2.8ex}{\small{{M4b}}}  & 6.2  & 7.7   & 31  & 7.7   \\
		\rule{0pt}{2.8ex}{\small{{M5}}}  & 1.8  & 1.8   & 8.0  & 1.8   \\
		\rule{0pt}{2.8ex}{\small{{M6a}}}  & .73  & .79   & 3.4  & .76   \\
		\rule{0pt}{2.8ex}{\small{{M6b}}}  & .36  & .37   & 3.3  & .42   \\
		\rule{0pt}{2.8ex}{\small{{M7a}}}  & 1.8  & 1.8   & 25  & 1.8   \\
		\rule{0pt}{2.8ex}{\small{{M7b}}}  & 10  & 10   & 48  & 10   \\
		\hline
	\end{tabular}
	\caption{The lowest possible scalar masses obtained in the scan for the various models, with $\chi_i^2<25$ and $m_{Z^\prime}<30$ TeV.}
	\label{tab:ScalarMasses}
\end{table}

\subsection{Collider bounds}

\begin{figure*}[ht!]
	\centering
	\begin{subfigure}[b]{0.5\textwidth}
		\includegraphics[width=\textwidth]{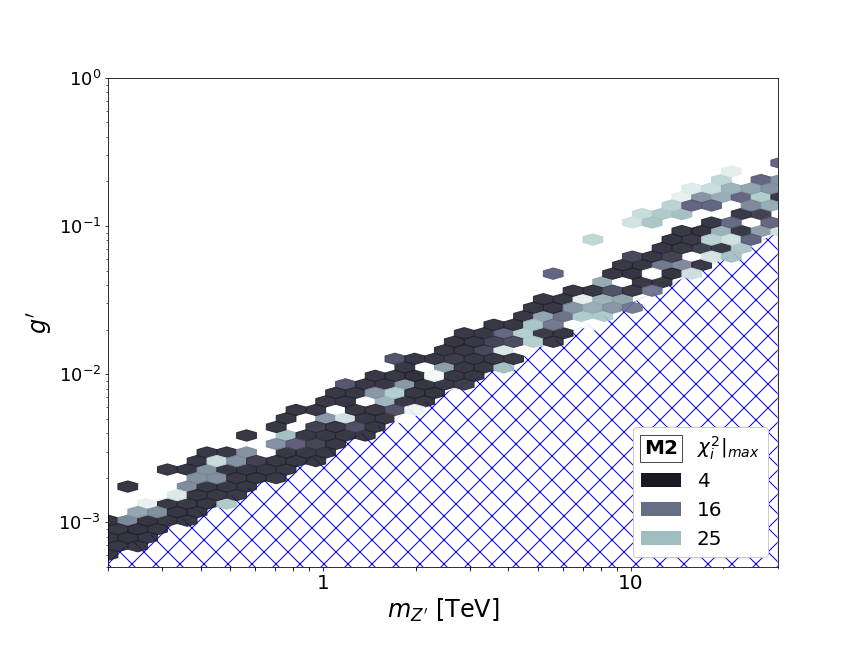}
	\end{subfigure}
	\hspace{-8mm}
	\begin{subfigure}[b]{0.5\textwidth}
		\includegraphics[width=\textwidth]{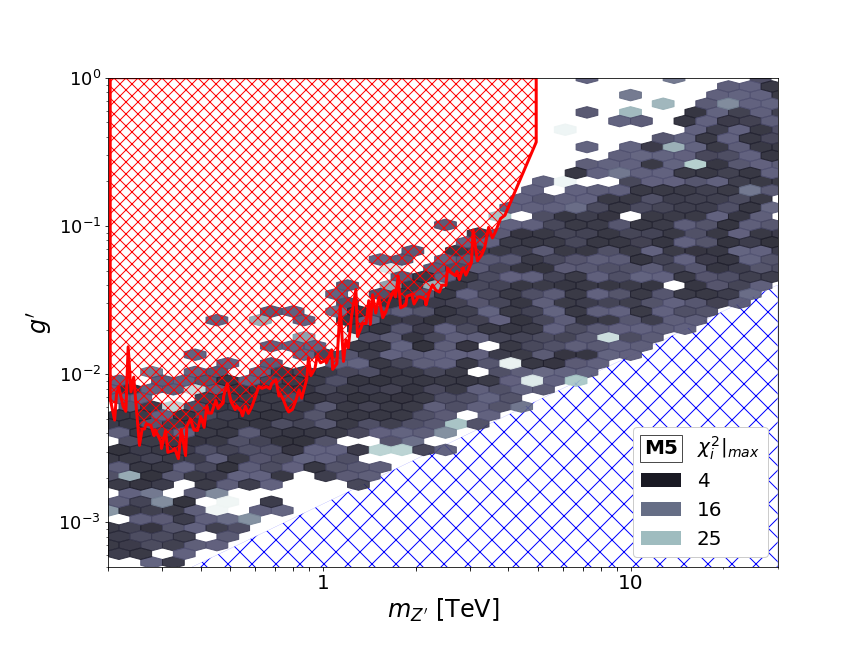}
	\end{subfigure}
		\begin{subfigure}[b]{0.5\textwidth}
		\includegraphics[width=\textwidth]{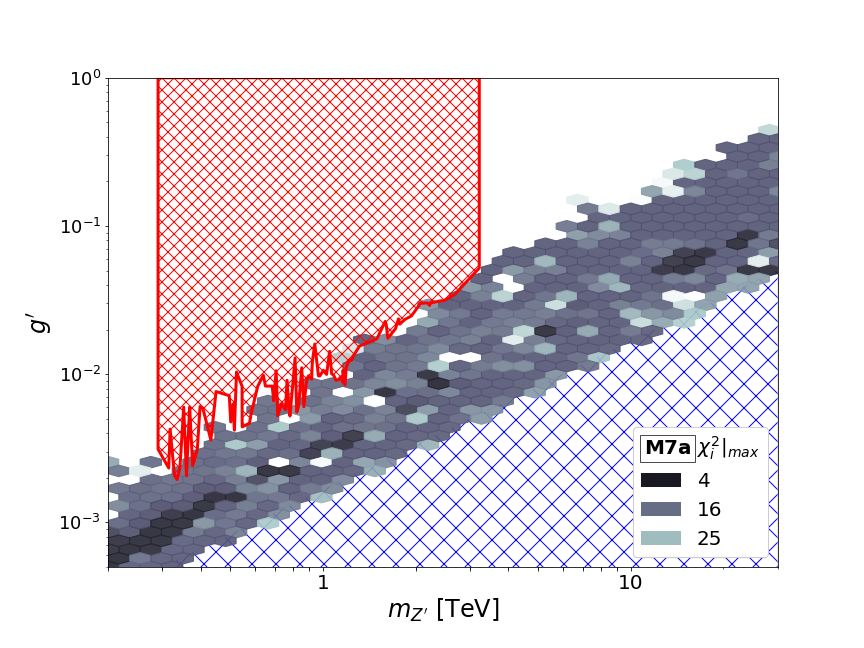}
	\end{subfigure}
	\hspace{-8mm}
	\begin{subfigure}[b]{0.5\textwidth}
		\includegraphics[width=\textwidth]{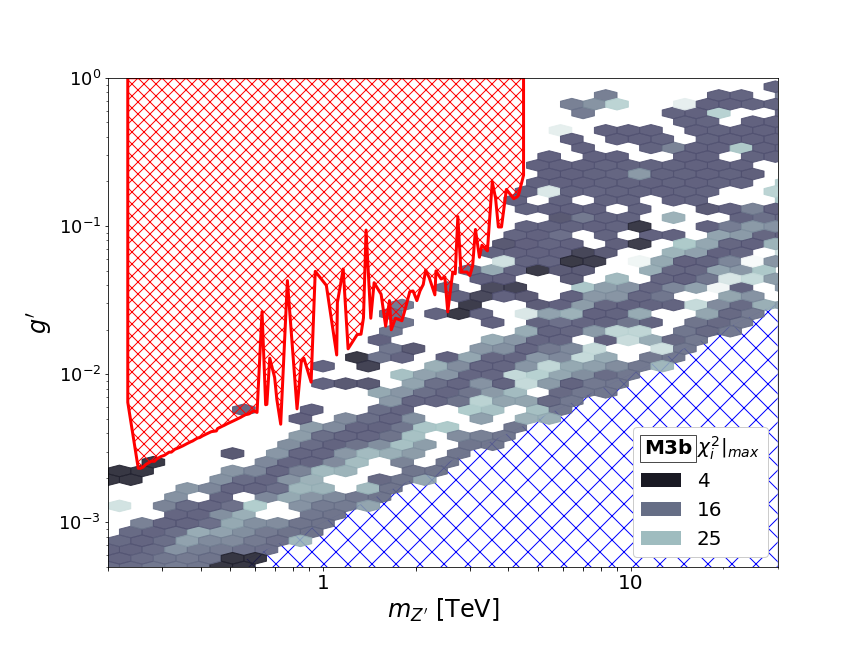}
	\end{subfigure}
	\caption{Collider constraints for models M2 (top left), M5 (top right), M7a (bottom left) and M3b (bottom right). The red, shaded area in the upper left corner indicates roughly the excluded regions, while the blue, shaded area in the lower right corner corresponds to an unscanned region of parameter space.} 
	\label{fig:ColliderM2}
\end{figure*}

In Fig.~\ref{fig:ColliderM2}, we plot $g^\prime$ as a function of the $Z'$ mass, for four of the models. Every parameter point excluded by collider constraints has been removed from the plot and instead replaced by a red, shaded area, for all larger values of $g^\prime$ with the same $Z^\prime$ mass. As such, the red, shaded area corresponds roughly to the area excluded by collider constraints. 
This area should not be seen as a forbidden region of the parameter space but more as a less favorable one, due to collider constraints. For example, in model M5, there are still several points inside this shaded red area that survived the collider bounds. These correspond to cases where the $Z^\prime$ coupling strength to light quarks is small, leading to a reduction in the production cross section. The blue, shaded area, with a larger grid, indicates any regions of parameter space not probed by our scans. 

From Eqs.~\eqref{eq:mass1} and \eqref{eq:mass2} we see that, in the case of a small mixing angle $\theta_M$ and a large hierarchy between the VEV of the scalar singlet and EW scale, the $Z'$ mass is approximately given by $m_{Z^\prime}\sim g^\prime v_S X_S$, explaining the mostly linear behavior of the plots. This also means that the tilt of the plot corresponds roughly to the inverse of the product of the $U(1)^\prime$ charge of the scalar singlet, and its VEV, i.e.~$(v_S X_S)^{-1}$. 

Given the charge assignments in Eq.~\eqref{eq:chargeAssignments}, the ratio $\Gamma_{Z^\prime}/m_{Z^\prime}$ was always below the $5\%$ level for all models, except for M5 ($7\%$) and M3b ($13\%$). When $g^\prime$ is large and $m_{Z^\prime}$ is small, we can have a Landau pole appearing at lower scales. A simple estimation, using the one-loop renormalization group equation for the evolution of $g^\prime$, gives, for the point $(g^\prime, m_{Z^\prime})=(1,10\,\text{TeV})$, a Landau pole at $10^9$ GeV for M3b and $10^{17}$ GeV for M5. Once $g^\prime$ is lower than $0.5$, all models have a Landau pole well above the grand unification scale.

Overall, models M2 and M5 show the most promising features. For model M2, the collider constraints do not even exclude any of the points. In comparison, the BGL model (not included in the plot) looks similar to model M5 but without reaching equally low $\chi^2$ values. 

\subsection{Ternary plots}

\begin{figure}[ht!]
    \centering
    \begin{subfigure}[b]{0.235\textwidth}
        \includegraphics[width=\textwidth]{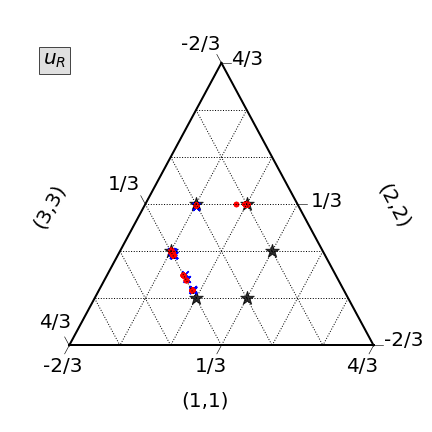}
    \end{subfigure}
    \begin{subfigure}[b]{0.235\textwidth}
        \includegraphics[width=\textwidth]{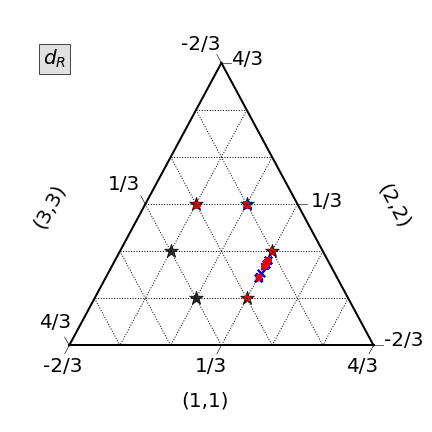}
    \end{subfigure}
    \caption{Preferred $U(1)'$ charges in the mass basis for model M2, with $x=1/3,\;y=0$ and $z=1$.} 
    \label{M2TriangleChargesHalf}
\end{figure}

\begin{figure}[ht!]
    \centering
\begin{subfigure}[b]{0.235\textwidth}
		\includegraphics[width=\textwidth]{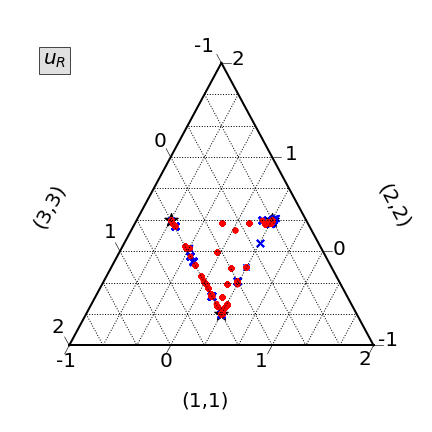}
	\end{subfigure}
	\begin{subfigure}[b]{0.235\textwidth}
		\includegraphics[width=\textwidth]{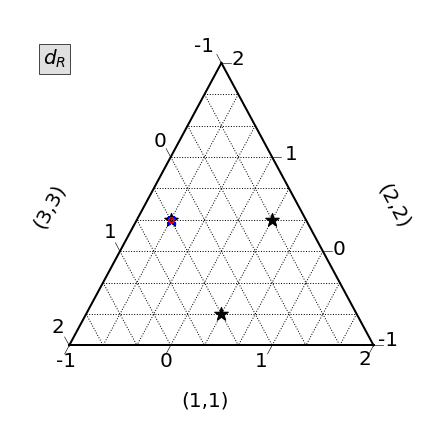}
	\end{subfigure}
  \caption{Preferred $U(1)'$ charges in the mass basis for model M3a, with $x=1/3,\;y=-2/3$ and $z=1$.} 
    \label{M3aTriangleChargesHalf}
\end{figure}

\begin{figure}[ht!]
    \centering
    	\begin{subfigure}[b]{0.235\textwidth}
		\includegraphics[width=\textwidth]{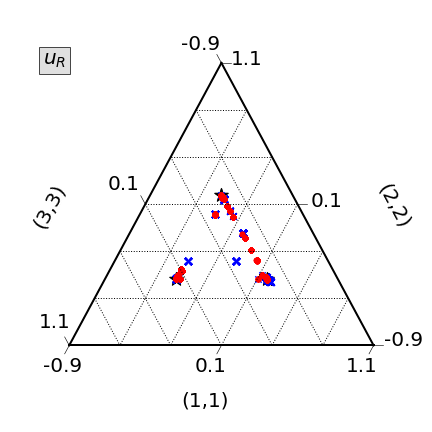}
	\end{subfigure}
	\begin{subfigure}[b]{0.235\textwidth}
		\includegraphics[width=\textwidth]{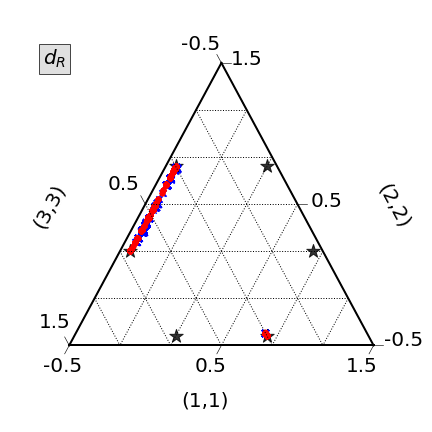}
	\end{subfigure}
	\caption{Preferred $U(1)'$ charges in the mass basis for model M7a, with $x=4/21$ and $y=-5/21$.} 
	\label{M7aTriangleChargesHalf}
\end{figure}

In Figs.~\ref{M2TriangleChargesHalf}-\ref{M7aTriangleChargesHalf}, we plot the preferred values for the $U(1)'$ charges in the mass eigenbasis of the gauge fields, $Q_{Z^\prime}^i/g^\prime$. The colored regions have a deviation below $5\sigma$ for all observables,\footnote{The corresponding plots for a lower deviation, e.g.~$3\sigma$, looks similar, but with fewer points and with less deviations from the black stars.} with the red points corresponding to a $Z-Z'$ mixing below $1\%$ and the blue points to a mixing above 1\%. As such, for the red points, the charge matrix in the mass and flavor eigenbases are approximately equal, i.e.

\begin{align}
\begin{split}
\frac{Q_{Z^\prime}^i}{g^\prime}\sim\frac{\hat{Q}_{Z^\prime}^i}{g^\prime}= \Xi_i.
\end{split}
\end{align}

In each plot, the black stars mark the $U(1)'$ charges in the flavor eigenbasis of the quarks. Hence, for any red parameter point positioned in the vicinity of a star, we have that $\Xi=U^\dagger \mathcal{X}U\sim \mathcal{X}$, i.e.~that the corresponding base changing matrix is close to unity. With this, the ternary plots can be used to visualize the amount of FCNCs that occur in a particular sector. 

With the CKM matrix given by {{$U_{\mathrm{uL}}^\dagger U_{\mathrm{dL}}$}}, the mixing matrices for the left-handed sectors are always rather close to unity. Hence, we will include only plots for the right-handed sectors. 

Take, for example, model M2, with $x=1/3$, $y=0$ and $z=1$. For these values, the charges for all quarks in the flavor eigenbasis are given by 

\begin{align}
\begin{split}
\mathcal{X}=\mathrm{diag}\{1/3,\;0,-1/3\},
\end{split}
\end{align}
such that there are six black stars in Fig.~\ref{M2TriangleChargesHalf}, namely, all permutations of $\{1/3,\;0,-1/3\}$. With the red points deviating by a small amount from the black stars, some off-diagonal contributions are allowed. 

For model M3a, there are two degenerate charges in all quark sectors. As such, there are three black stars in Fig.~\ref{M3aTriangleChargesHalf} rather than six. Here, the right-handed up sector shows a significant deviation, while, for the down sector, the parameter points are practically on top of one of the black stars. Similar features can be found in model M3b but with the FCNCs being slightly subdued in the up sector and with some off-diagonal elements also in the down sector. 

In Fig.~\ref{M7aTriangleChargesHalf}, we have the preferred $U(1)^\prime$ charges for model M7a. Here, there are sizable off-diagonal elements in both the right-handed sectors. For model M7b, the base-changing matrices are slightly closer to unity.

Overall, we see that the ternary plots are dominated by red parameter points; i.e.~there is a strong preference toward a small $Z-Z^\prime$ mixing.

\subsection{Flavor anomalies}\label{subsec:flavorAno}

\begin{table*}
\begin{tabular}{lll lll lll lll lll} 
\hline\hline 
\\
&{\small{\bf{M2}}}&& {\small{\bf{M3a,b}}}&&{\small{\bf{M4a}}}&&{\small{\bf{M4b}}}&&{\small{\bf{M5}}}&& {\small{\bf{M7a}}}&&{\small{\bf{M7b}}}
 \\[2mm]
\hline
\\
 \rule{0pt}{2.8ex}{\small{$(1,2,3)\hspace{5mm}$}}  & \cellcolor{gray!20} $0\;(0)\hspace{10mm}$ 
  && 
  \cellcolor{gray!20} $0\;(0)\hspace{10mm}$ 
 &&
\cellcolor{gray!20} $0\;(0)\hspace{10mm}$
 && 
 $0\;\left(-\dfrac{13}{22}\right)\hspace{10mm}$
 && 
 \cellcolor{gray!20} $0\;(0)\hspace{10mm}$ 
  &&
 $\dfrac{1}{3}^*\;(\infty)^*\hspace{10mm}$   
 && 
 $0\;\left(-\dfrac{11}{2}\right)$       
 \\[4mm]
 \rule{0pt}{2.8ex}{\small{$(1,3,2)$}}  & $0\;(0)$   
  && 
   $0\;(0)$
 && 
 $\dfrac{3}{23}\;(0)$ 
 &&
 $\dfrac{1}{3}^*\;(\infty)^*$   
 && 
 $\dfrac{3}{5}\;(0)$   
 && 
 $-3\;(-10)$ 
 &&
 $\dfrac{1}{3}\;(\infty)^*$   
 \\[4mm]
 \rule{0pt}{2.8ex}{\small{$(2,1,3)$}} &  $\dfrac{1}{3}^*\;(\infty)^*$  
  &&   
 $\dfrac{1}{3}^*\;(\infty)^*$      
 &&
  $\dfrac{1}{3}^*\;(\infty)^*$    
  &&
  $-\dfrac{3}{13}\;\left(-\dfrac{23}{13}\right)$ 
  &&
  $\dfrac{1}{3}^*\;(\infty)^*$
 &&
 \cellcolor{gray!20} $0\;(0)$
  &&
  $\dfrac{3}{11}\;\left(-\dfrac{2}{11}\right)$      
  \\[4mm]
 \rule{0pt}{2.8ex}{\small{$(2,3,1)$}} & $0\;(-1)$ 
  && 
     $0\;(-1)$
 &&
$\dfrac{3}{23}\;\left(-\dfrac{14}{23}\right)$ 
  && 
 $\dfrac{1}{3}^*\;(\infty)^*$
 &&
 $\dfrac{3}{5}\;\left(\dfrac{4}{5}\right)$ 
 &&
$-3\;(0)$
  && 
 $\dfrac{1}{3}^*\;(\infty)^*$
 \\[4mm]
 \rule{0pt}{2.8ex}{\small{$(3,1,2)$}}& $\dfrac{1}{3}\;(\infty)^*$
  && 
   $\dfrac{1}{3}^*\;(\infty)^*$
 &&
 $\dfrac{1}{3}^*\;(\infty)^*$ 
 &&
 $-\dfrac{3}{13}\;(0)$ 
 &&
 $\dfrac{1}{3}^*\;(\infty)^*$ 
 &&
 $0\;\left(-\dfrac{1}{10}\right)$
 &&
 $\dfrac{3}{11}\;(0)$
 \\[4mm]
 \rule{0pt}{2.8ex}{\small{$(3,2,1)$}}  & $0\;(-1)$
  && 
   $0\;(-1)$
 &&
 $0\;\left(-\dfrac{23}{14}\right)$
 && 
 \cellcolor{gray!20} $0\;(0)$ 
 &&
 $0\;\left(\dfrac{5}{4}\right)$   
 &&
 $\dfrac{1}{3}^*\;(\infty)^*$ 
 && 
 \cellcolor{gray!20} $0\;(0)$ \\[4mm]
\hline\hline
\end{tabular}
\caption{Values for ${C_{10}^{\mathrm{NP}\mu(\prime)}}/{C_9^{\mathrm{NP}\mu(\prime)}}\;\left({C_{9}^{\mathrm{NP}e(\prime)}}/{C_9^{\mathrm{NP}\mu(\prime)}}\right)$. Here, the asterisk indicates entries for which $C_9^{\mathrm{NP}\mu(\prime)}$ is zero, while shaded cells corresponds to the flavor configurations used in the scans.}
\label{tab:RKpart2}
\end{table*}

The values for the ratios of $C_{9,10}^{\mathrm{NP}(\prime)}$, with the charge assignments in Eq.~\eqref{eq:chargeAssignments}, are shown in Table \ref{tab:RKpart2}. Given these charges and the identification of the electron as the charged lepton with no $U(1)^\prime$ charge, the Wilson coefficients $C_{9,10}^{\mathrm{NP}e(\prime)}$ are automatically zero for all models. It turns out that, for this choice, there is always a scenario where the muon couples vectorially to the $Z^\prime$, leaving $C_9^{\mathrm{NP}\mu (\prime)}$ as the only nonzero Wilson coefficient. In Table \ref{tab:RKpart2}, these cases are indicated by gray, shaded cells, which are also the scenarios used in the scan. Note that, for models M5 and M6b (and M1), the unprimed Wilson coefficients are zero, as there are no FCNCs in the left-handed down sector. Similarly, the primed Wilson coefficients are zero for models M6a and M6b (and M1), as they have no FCNCs in the right-handed down sector. 

\begin{figure}[ht!]
	\centering
	\begin{subfigure}[b]{0.5\textwidth}
		\includegraphics[width=\textwidth]{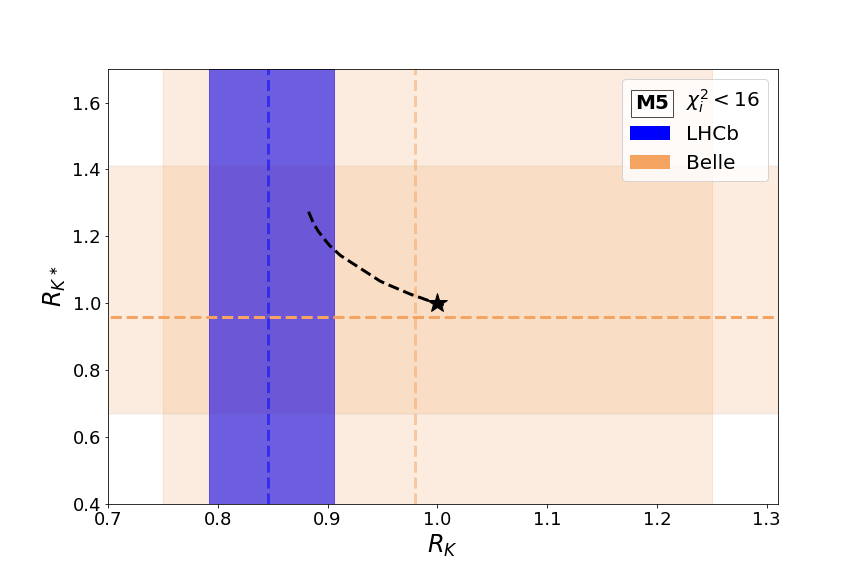}
	\end{subfigure}
	\caption{$R_K^*$ as a function of $R_K$ for model M5, with the blue area indicating the $1\sigma$ LHCb bound, the orange area the Belle one, and where the SM value is indicated by a star.} 
	\label{fig:RKRKstar}
\end{figure}

For most models, the predictions for $R_K$ and $R_{K^\ast}$ were just the SM ones. However, for model M5, the presence of $C_9^{\mathrm{NP}\mu \prime}$ was enough to have sizable deviations from the SM prediction. In Fig.~\ref{fig:RKRKstar}, we plot the ratio for $R_K^*,R_K$ for model M5, together with the $1\sigma$ bounds for the LHCb and Belle measurements, as specified in Sec.~\ref{subsec:meson}. Here we see that there are parameter points that can account for the current anomaly. Note that, in the plot, we are showing the values only in the region of interest for the anomaly - there exist plenty of points in agreement with the SM prediction and also with a larger value for $R_K$ than 1.

We have also looked into the predictions of these models for the ratios $R^\nu_{K,K^\ast}$ as defined in Ref.~\cite{Altmannshofer:2009ma}. Most model predictions were compatible with the SM results. However, models M4a and M5 allowed some significant deviations. For model M4a $(R_{K}^\nu,R_{K^\ast}^\nu)$ varied from $(0.8,1.1)$ to $(1.3,0.85)$ linearly. In model M5, since there are only right-handed FCNCs, the two ratios are equal and were in the range $[0.95,1.45]$.

\section{Conclusions}\label{s:con}


This work has two central results. One is the classification of all anomaly-free 2HDMs with a gauged flavor symmetry, and the other is the analysis of their phenomenological consistency. Together, the two components can be used to identify the current research gap for this type of 2HDM and work as a starting point for further studies. 

Out of the 116 model candidates corresponding to continuous symmetries in Ref.~\cite{Ferreira:2010ir}, only 11 models survived after imposing anomaly cancellation. Six of these models were studied here for the first time. In the phenomenological study, each model was subjected to a range of observational constraints, and while there were good parameter points for all 11 models, we wish to give extra attention to models M2 and M5. For these two models, one could easily find extended regions of parameter space with a deviation below $3\sigma$, in addition to model M5 being able to accommodate the $R_K,R_K^\ast$ anomaly with only the presence of $C_9^{\mathrm{NP}\mu \prime}$. As such, they deserve further consideration.

\begin{acknowledgments}

The authors thank Johannes Bellm for enlightening discussions regarding collider bounds. 
H.S. also thanks Jo\~ao P. Silva for the careful reading and valuable comments on the manuscript. 
This work is supported in part by the Swedish Research Council,
Contract No.~2016-05996, and in part by the European Research
Council (ERC) under the European Union's Horizon 2020
research and innovation program, Grant Agreement No.~668679.

\end{acknowledgments}

\begin{appendix}

\section{Scalar potential}\label{app:scalar}

The scalar potential is given by $V=V_0 + \{V_1+\mathrm{H.c.}\}$, where $V_0$ is the phase-insensitive part

\begin{align}
\begin{split}
V_0&=\sum_i \left(m_i^2\left|\Phi_i\right|^2+\lambda_{i}\left|\Phi_i\right|^4\right)\\
&\;\;\;\; + \lambda_{12}\left|\Phi_1\right|^2\left|\Phi_2\right|^2 + \lambda^\prime_{12}\left|\Phi_1^\dagger \Phi_2\right|^2
\\
&\;\;\;\;+m_S^2\left|S\right|^2+\lambda_S\left|S\right|^4 + \sum_i \lambda_{Si}\left|\Phi_i\right|^2\left|S\right|^2,
\end{split}
\end{align}
with $i=1,2$, and $V_1$ the phase-sensitive part

\begin{align}
\hspace{-2mm}V_1\hspace{-0.5mm}=\hspace{-0.5mm}\begin{cases}
\hspace{0mm}a_{1(2)}\Phi_1^\dagger \Phi_2S^{(*)}\hspace{1.0mm}\mathrm{for}\hspace{0.5mm}X_S\hspace{-0.5mm}=\hspace{-0.5mm}\varpm \left( X_{\Phi_1}- X_{\Phi_2}\right)
\\[1.2mm]
\hspace{0mm}a_{3(4)}\Phi_1^\dagger \Phi_2{S^{(*)}}^2\hspace{0mm}\mathrm{for}\hspace{0.5mm}X_S\hspace{-0.5mm}=\hspace{-0.5mm}\varpm( X_{\Phi_1}- X_{\Phi_2})/2
\end{cases}
\end{align}
where $ X_{\Phi_{1,2}}$ are the $U(1)^\prime$ charges of the Higgs doublets and $X_S$ the $U(1)^\prime$ charge of the scalar singlet. 

Note that only one of the four parameters in $V_1$ can be nonzero at a time. As such, we can always choose all parameters to be real by a rephasing of the scalar singlet $S$, resulting in the scalar potential being $CP$ conserving. This, in turn, also inhibits the possibility of spontaneous $CP$ violation \cite{Branco:2015bfb}. 

Below, we present the complex tadpole equations. In expressions containing both $a_{1(2)}$ and $a_{3(4)}$ simultaneously, it is always implicit that only one of them can be present for each scalar potential variation. We then have

\begin{align}
\begin{split}
0\equiv&\left(M_\pm^2\right)_{ij} v_j\,,\\
0\equiv&\left(m_S^2+\lambda_S|v_S|^2+\frac{1}{2}(\lambda_{S1}v_1^2+\lambda_{S2}v_2^2)\right) v_S e^{i\alpha_S} \\
&\quad+ v_1v_2\left(\frac{a_{1(2)}}{\sqrt{2}}+a_{3(4)}v_Se^{-i\alpha_S}\right)\,,
\end{split}
\end{align}
with $M_\pm^2$ the Hermitian squared mass matrix for the charged scalars, defined as $\phi_a^- \left(M_{\pm}^2\right)_{ab} \phi^+_b$ and given by

\begin{align}
\begin{split}
\left(M_\pm^2\right)_{ii}=&m_i^2 +\lambda_i v_i^2+\frac{\lambda_{Si}^\prime}{2}v_S^2\\
&+\frac{\lambda_{12}}{2}(v_2^2\delta_{i1}+v_1^2\delta_{i2})\,,\\
\left(M_\pm^2\right)_{12}=&\frac{\lambda^\prime_{12}}{2}v_2v_1+\frac{1}{\sqrt{2}}a_{1(2)}v_Se^{\varpm i\alpha_S}\\
&+\dfrac{1}{2}a_{3(4)}v_S^{2} e^{\varpm i2\alpha_S}\,.
\end{split}
\end{align}
We can take the imaginary part of the tadpole equations and explicitly see that the phase of the singlet VEV can be only 0 or a multiple of $\pi$, for $a_{1(2)}$, and multiple of $\pi/2$, for $a_{3(4)}$. From this point onwards, all expressions take the singlet phase to be zero. For the neutral scalars, the mass term is defined as

\begin{align}
\begin{split}
\frac{1}{2}\varphi^{0 T} M_0^2 \varphi^0\quad \text{with}\quad
\varphi^0 = 
\begin{pmatrix}
R_a\\
I_a\\
\rho\\
\eta
\end{pmatrix}\,,
\end{split}
\end{align}
following the notation used in Eq.~\eqref{eq:Scalars}. Since $CP$ is conserved, there is no mixing between the $CP$-even and $CP$-odd scalar components, leading to the simplified symmetric neutral mass matrix

\begin{equation}
M_0^2=\begin{pmatrix}
\left.M_{Re}^2\right|^{\phi\phi}&0&\left.M_{Re}^2\right|^{\phi S}&0\\
&\left.M_{Im}^2\right|^{\phi\phi}&0&\left.M_{Im}^2\right|^{\phi S}\\
&&\left.M_{Re}^2\right|^{SS}&0\\
&&&\left.M_{Im}^2\right|^{SS}
\end{pmatrix}\,, 
\end{equation}
with the different entries explicitly given by

\begin{align*}
\begin{split}
\left.M_{Re}^2\right|^{\phi\phi}=&\;
M_\pm^2 +
\begin{pmatrix}
2\lambda_1 v_1^2+\frac{\lambda_{12}^\prime}{2}v_2^2&(\lambda_{12}+\frac{\lambda_{12}^\prime}{2})v_1 v_2\\
(\lambda_{12}+\frac{\lambda_{12}^\prime}{2})v_1 v_2&2\lambda_2 v_2^2+\frac{\lambda_{12}^\prime}{2}v_1^2
\end{pmatrix}\,,\\
\left.M_{Im}^2\right|^{\phi\phi}=&\;
M_\pm^2+\frac{1}{2}\lambda_{12}^\prime
\begin{pmatrix}
 v_2^2&-v_1 v_2\\
-v_1 v_2&v_1^2
\end{pmatrix}\,,\\
\left.M_{Re}^2\right|^{SS}=&\;m_S^2 +3 \lambda_S v_S^2+\frac{1}{2}(\lambda_{S1}v_1^2+\lambda_{S2}v_2^2)\\
&+a_{3(4)}v_1v_2\,,\\
\left.M_{Im}^2\right|^{SS}=&\;m_S^2 +\lambda_S v_S^2+\frac{1}{2}(\lambda_{S1}v_1^2+\lambda_{S2}v_2^2)\\
&-a_{3(4)}v_1v_2\,,
\end{split}
\end{align*}

\pagebreak

\begin{align*}
\begin{split}
\left.M_{Re}^2\right|^{\phi S}=&
\begin{pmatrix}
\lambda_{S1}v_1 v_S + \frac{1}{\sqrt{2}}a_{1(2)}v_2+a_{3(4)}v_Sv_2\\
\lambda_{S2}v_2 v_S + \frac{1}{\sqrt{2}}a_{1(2)}v_1+a_{3(4)}v_Sv_1
\end{pmatrix}\,,
\\
\left.M_{Im}^2\right|^{\phi S}=&
\begin{pmatrix}
\frac{1}{\sqrt{2}}a_{1(2)}v_2+a_{3(4)}v_Sv_2\\
-\frac{1}{\sqrt{2}}a_{1(2)}v_1-a_{3(4)}v_Sv_1
\end{pmatrix}\,.
\end{split}
\end{align*}

\section{Anomaly conditions}\label{app:anomaliesCond}

There are six anomaly constraints involving $U(1)^\prime$ which does not cancel trivially, namely

\begin{align}
\label{eq:ano2}
\begin{split}
A_{111'}&=\sum_{i=1}^3\left( X_{q_i}+3X_{\ell_i}-8X_{u_i}-2X_{d_i}-6X_{e_i}\right),
\\
A_{11'1'}&=\sum_{i=1}^3\left(X_{q_i}^2-X_{\ell_i}^2-2X_{u_i}^2+X_{d_i}^2+X_{e_i}^2\right),
\\
A_{1'1'1'}&=\sum_{i=1}^3\left(6X_{q_i}^3+2X_{\ell_i}^3-3X_{u_i}^3-3X_{d_i}^3-X_{e_i}^3\right),
\\
A_{221'}&=\sum_{i=1}^3\left(3X_{q_i}+X_{\ell_i}\right)
\\
A_{331'}&=\sum_{i=1}^3\left(2X_{q_i}-X_{u_i}-X_{d_i}\right),
\\
A_{gg1'}&=\sum_{i=1}^3\left(6X_{q_i}+2X_{\ell_i}-3X_{u_i}-3X_{d_i}-X_{e_i}\right),
\end{split}
\end{align}
where $A_{\mathrm{XYZ}}\propto\mathrm{Tr}\left(\{T_\mathrm{X},T_\mathrm{Y}\}T_\mathrm{Z}\right)$, with $T_\mathrm{X}$ being a generator in the fundamental representation for the gauge group $\mathrm{X}$, and where $A_{gg1'}$ is the gravitational anomaly.

\end{appendix}

\bibliography{bib}

\end{document}